\documentclass[12pt]{article}
\usepackage{graphicx}
\usepackage{epstopdf}
\usepackage[body={17.5cm, 22cm},right=2cm]{geometry}
\usepackage{amssymb}

\newcommand\nn{\nonumber}
\newcommand\ee{\end{equation}}
\newcommand\be{\begin{equation}}
\newcommand\eea{\end{eqnarray}}
\newcommand\bea{\begin{eqnarray}}

\def\la{\langle}
\def\ra{\rangle}
\def\r{\right}
\def\l{\left}

\def\beq{\begin{equation}}
\def\eeq{\end{equation}}
\def\d{\partial}

\def\Oh{{\sqrt{\Omega}}}
\def\eh{{\sqrt{\epsilon}}}
\def\de{{\partial}}

\begin{document}
\setcounter{page}{0}
\thispagestyle{empty}

\begin{titlepage}

~\vspace{1cm}
\begin{center}

{\LARGE \bf The Volume of the Universe after Inflation \\[.4cm] and de~Sitter Entropy}

\vspace{1.2cm}

{\large \bf
Sergei Dubovsky$^{a,b}$,
Leonardo Senatore$^{c,d,e}$, and 
Giovanni Villadoro$^{f}$}
\\
\vspace{.45cm}
{\normalsize { \sl $^{a}$ Department of Physics, Stanford University, Stanford, CA 94305 }}\\ 

\vspace{.3cm}
{\normalsize { \sl $^{b}$
Institute for Nuclear Research of the Russian Academy of Sciences, \\
        60th October Anniversary Prospect, 7a, 117312 Moscow, Russia}}

\vspace{.3cm}
{\normalsize { \sl $^{c}$ School of Natural Sciences, Institute for Advanced Study, \\Olden Lane, Princeton, NJ 08540, USA}}\\

\vspace{.3cm}
{\normalsize { \sl $^{d}$ Jefferson Physical Laboratory, Harvard University, Cambridge, MA 02138, USA}}\\

\vspace{.3cm}
{\normalsize {\sl $^{\rm e}$  
Center for Astrophysics, Harvard University, Cambridge, MA 02138, USA}}

\vspace{.3cm}
{\normalsize { \sl $^{f}$ CERN, Theory Division, CH-1211 Geneva 23, Switzerland}}
\end{center}
\vspace{.8cm}
\begin{abstract}
We calculate the probability distribution for the volume of the Universe
after slow-roll inflation both in the eternal and in the non-eternal regime. Far from the eternal regime the probability
distribution for the number of $e$-foldings, defined as one third of the logarithm of the volume, 
is sharply peaked around the number of $e$-foldings of the
classical inflaton trajectory. 
At the transition to the eternal regime  this probability is still peaked (with the width of order one $e$-folding) 
around the  average, which gets twice larger
at the transition point. As one enters the eternal
regime the probability for the volume to be finite rapidly becomes exponentially small. 
In addition to developing techniques to study eternal inflation, 
our results allow us to establish the quantum generalization of a recently proposed bound 
on the number of $e$-foldings in the non-eternal regime: the probability for slow-roll inflation 
to produce a finite volume larger than $e^{S_{dS}/2}$, where $S_{dS}$ is the de~Sitter
entropy at the end of the inflationary stage, is smaller than the uncertainty due to non-perturbative quantum gravity
effects. The existence of such a bound provides a consistency check for the idea of de Sitter 
complementarity.
\end{abstract}

\end{titlepage}



\section{Introduction}
The Universe is accelerating today~\cite{SSTC,SCPC,WMAP}, and it is extremely likely that 
it was experiencing a period of accelerated expansion (inflation) back in the past~\cite{WMAP,Guth:1980zm,Linde:1981mu,Albrecht:1982wi}.
In both cases the pressure to density ratio is very close to $-1$ and the local geometry is very close to 
that of de~Sitter (dS) space.  

It is plausible that both these periods of inflation are eternal, {\it i.e.}  some space-time regions
keep inflating forever.  
Indeed the most economical explanation for the cosmic acceleration observed now
is that we are stuck in a metastable vacuum \cite{Guth:1982pn}, and this results in eternal inflation unless the vacuum decay rate  
$\Gamma$ is faster than the expansion rate of the Universe, $\Gamma\gtrsim H_0$~\footnote{
This latter possibility appears rather unlikely and fine-tuned given the non-perturbative nature of the decay,
although it may receive support from future particle physics data \cite{ArkaniHamed:2008ym}.}. 
Also there is strong evidence that in the past we underwent a phase of inflation driven by a rolling scalar field,
which could have been preceeded by a period of eternal inflation, as predicted in many field theoretical models~\cite{Vilenkin:1983xq,Linde:1986fd,Linde:1986fc}.  
Further, the current picture of the string landscape~\cite{Douglas:2006es} suggests that the observed part of the Universe was created 
as a result of tunneling from some higher-scale eternally-inflating  vacuum \cite{Bousso:2000xa}. 
All these arguments make the study of eternal inflation very important.

Moreover eternal inflation provides a natural framework to implement Weinberg's solution of the cosmological constant problem~\cite{Weinberg:1987dv},
which is the most plausible so far in spite of the many efforts made to find an alternative.
According to this solution the choice of the vacuum is  
made, at least partially, by anthropic reasons such as the requirement that structures were able to form in our Universe.
These arguments raise the notoriously difficult and puzzling question of making predictions in an eternally inflating Universe.

On a purely practical side we possess a well developed machinery of quantum field theory in curved space-time
that proved to be very successful in calculating properties of the primordial density perturbations with many fine
details in the case of non-eternal inflation (see e.g. \cite{Maldacena:2002vr}). The applications of these techniques to the case  of eternal inflation 
is instead much more challenging, as in this latter regime the size of the quantum fluctuations is large, 
a non-perturbative treatment is required and calculations become in general much more difficult (see e.g. \cite{Creminelli:2008es} for a recent discussion). 
On the other hand, without a clear understanding of the eternally inflating geometry, it might be hopeless
to solve the issues raised by eternal inflation such as the measure problem in the landscape; this is why
we find it very important to make explicit and precise calculations in this regime. 

On a more theoretical side, dS space appears to share many properties with the black hole
geometry---most importantly, in both space-times the most natural sets of observers (the asymptotic observers in 
the black hole case and the comoving ones in the accelerating Universe) see a gravitational horizon with 
the associated thermodynamic properties, such as Hawking temperature
and Bekenstein entropy, the latter, in the case of de~Sitter, being equal to \cite{Gibbons:1977mu}
\[
S_{dS}=\pi{M_{\rm Pl}^2\over H^2}\;,
\]
where $M_{\rm Pl}^2\equiv 1/G_N$ with $G_N$ the Newton constant.
 A finite entropy suggests that the system is described by a finite number of degrees of freedom, or more formally,
 the Hilbert space describing the system at the quantum level has a finite dimensionality equal to $e^{S_{dS}}$.
 It is widely believed that both black holes and de~Sitter space always arise as a subsector of a larger theory 
 with an infinite dimensional Hilbert space.
 This is obvious for a black hole in asymptotically Minkowski or adS space-times, but less clear
 for dS space. Indeed, in principle one could imagine a quantum gravity theory with, for example, 
 a single positive energy vacuum. However, this is
 not the case in the known string theory landscape~\cite{Kachru:2003aw} 
 and there are general arguments strongly suggesting that dS vacua are always metastable with respect
 to decay to either the Minkowski or the adS minima of the potential~\cite{Goheer:2002vf}. 
 This will be the point of view adopted in the current paper and by dimensionality of the Hilbert space describing
 black hole or dS space we always understand the dimensionality of the corresponding subsector in a larger, 
 most likely infinite-dimensional, Hilbert space (although, it is worth noting 
 that an alternative line of thought is also being pursued \cite{Banks:2005bm}). 
 
Taking seriously the similarity between the causal structures of de~Sitter and Schwarzschild
causes a serious doubt on the validity of the global semiclassical picture of the eternally inflating Universe\footnote{This idea is being pursued by a number of authors, see {\it e.g.} \cite{Bousso:2006ge}.  Our
discussion  mainly follows that of  \cite{ArkaniHamed:2007ky}.}.
Indeed, the remarkable fact about black holes is that the global
 effective field theory description of the space-time claiming
 to describe both the interior and the exterior of the horizon eventually breaks down~(see e.g. \cite{Lowe:1995ac}). 
 
\begin{figure}[t]
\begin{center}
\includegraphics[height=4cm]{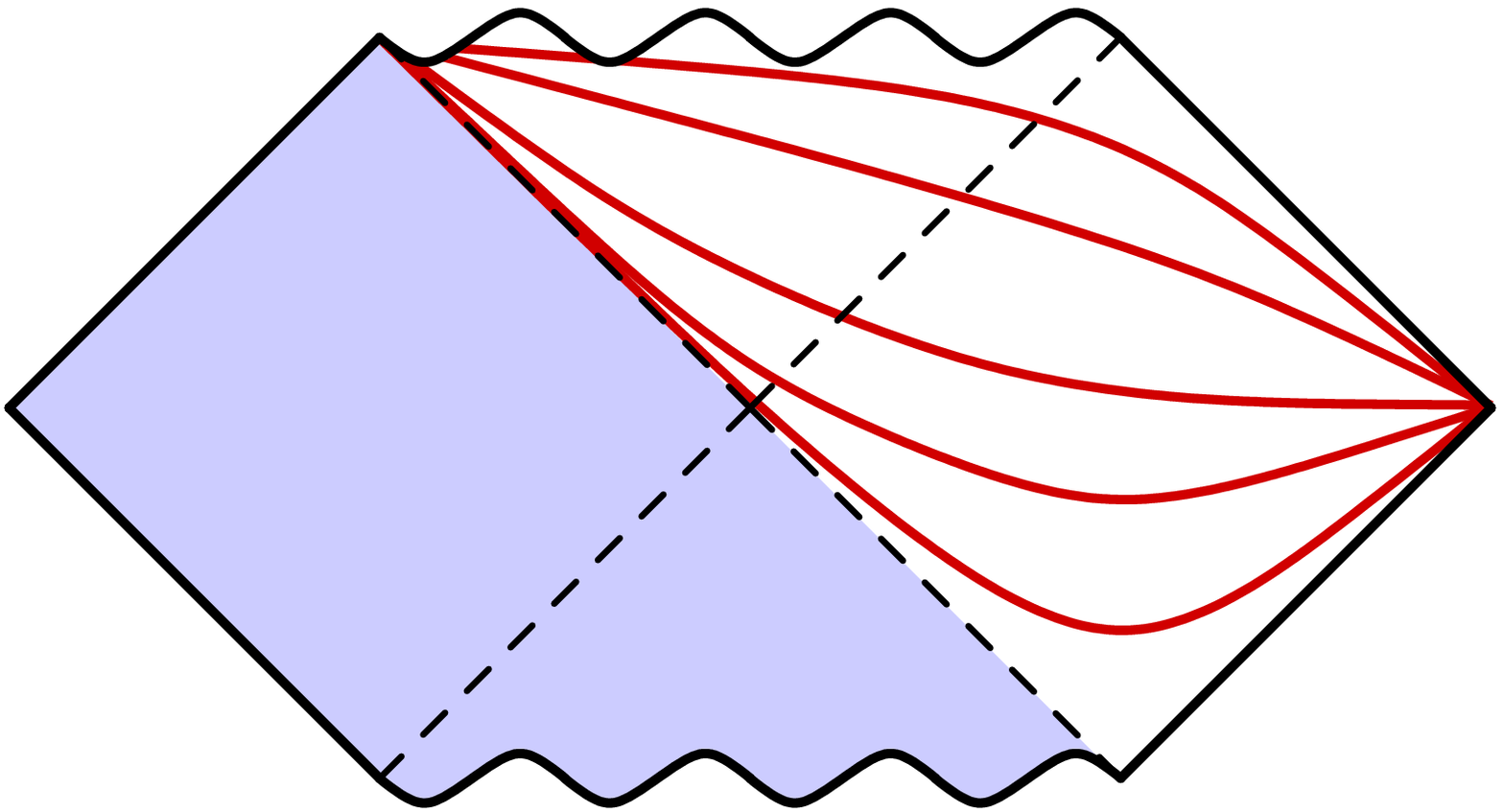} \hspace{2cm}
\includegraphics[height=4cm]{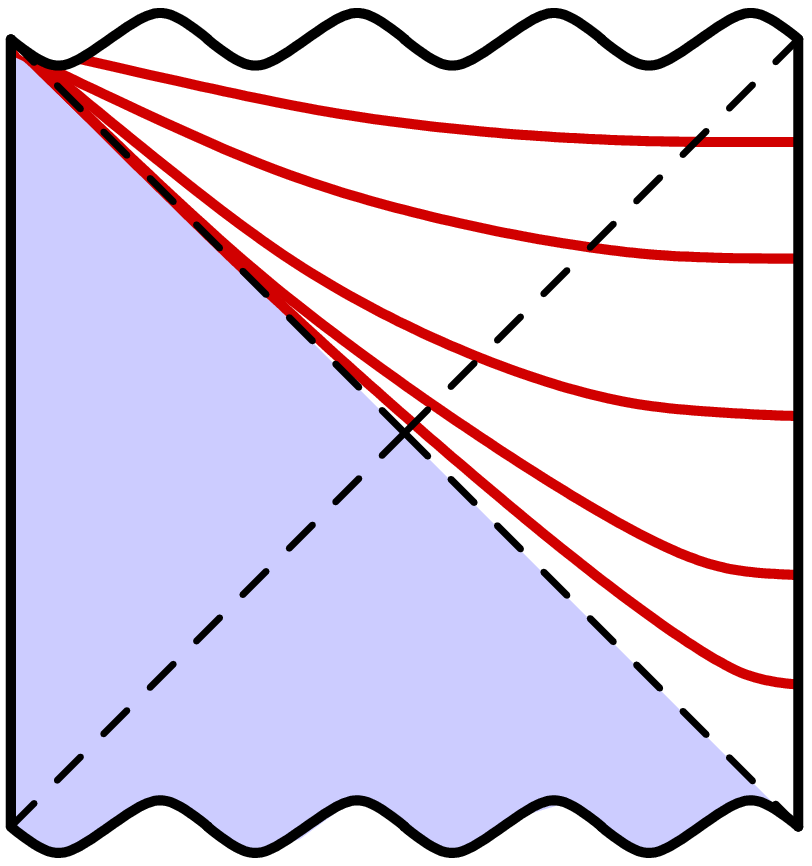}
\caption{\label{fig:penrose} \small \it Set of space-like slices covering both the exterior and the interior of: 
a) a black hole (left) and b) an eternal inflating universe (right).}
\end{center}
\end{figure}

 More concretely, if one considers the set of space-like slices covering both the exterior and the interior of a black hole, as shown in 
 fig.~\ref{fig:penrose}a, and insists that the field theory description is valid on this set of slices, one comes to the conclusion that the information is lost in the course of the black hole evaporation~\cite{Hawking:1976ra}. This conclusion was proven to be wrong \cite{Maldacena:2001kr,Hawking:2005kf} by the adS/CFT
 correspondence  that provides a description of the system involving black holes (gravity in the bulk adS space) in terms of an unitary boundary CFT.
Consequently, one is forced to conclude that the global effective field theory description 
 breaks down at the time scale of order the evaporation time, $t\sim R_s^3/G_N$, where $R_s$ is the Schwarzschild radius and $G_N$ the Newton constant (for a recent review see for example \cite{ArkaniHamed:2007ky}).
 After this time the information about the inside observer gets reprocessed in the evaporating Hawking
 quanta, and by insisting on a simultaneous local description of the exterior and the interior on longer time-scales one would run
 into a contradiction with the ``no quantum xerox" principle (or equivalently with the linearity of quantum mechanics) \cite{'tHooft:1990fr,Susskind:1993if}. 
 This conclusion is really surprising given that one can always choose a set of slices that avoid the region close to the singularity, 
 so that naively one would expect the effective field theory to hold.

 Given the similarity between the causal structures of de~Sitter space and that of Schwarzschild geometry,
one may suggest that also in dS space the global description in terms of a similar set of slices (shown in fig.~\ref{fig:penrose}b)
eventually breaks down.  Note that these slices are just the conventional FRW slices that are commonly used to describe
the inflationary Universe. Going on with the analogy with the black holes one expects this breakdown to happen at a time-scale
of order $t\sim H^{-3}/G_N$ or, equivalently, after a period of order $S_{dS}$ $e$-foldings.  One may expect that space-time events outside
the region containing $e^{S_{dS}}$ Hubble patches get encoded in de~Sitter fluctuations,  similarly to how the information inside the black hole gets
released in the Hawking quanta after the evaporation time.

Note that, unlike in the black hole case, we are not running into any paradox if this does not happen, therefore it may well be
that pushing the analogy this far is too naive. However, if one takes it seriously there is an immediate test to pass.
Indeed, in the black hole case, due to the presence of a curvature singularity, it is impossible to read the information in the Hawking quanta and then jump
into the black hole and read the same information again. 
Similarly,  it should not be possible to get the same information twice in the de~Sitter case as well. 

However, if inflation
could last for an arbitrarily long time without becoming eternal, an observer would be able to first read the information in the dS fluctuations
and later, as more and more space-time becomes visible after inflation, he would
 be able to access it directly. Consequently, for the above ideas about de~Sitter complementarity to be consistent,
 there should be a limit on how long inflation can last without becoming
 eternal \cite{ArkaniHamed:2007ky},
 \be
 \label{vague_limit}
 3N\leq cS_{dS}\;,
 \ee
 where $N$ is the number of $e$-foldings (defined as one third of the logarithm of the total volume after inflation) and $c$ is a coefficient of order one. On the other hand, no limit on the number of $e$-foldings is required on those realization that are eternal, because in those cases the observer is not able to access all the volume after reheating.
 
 It was proven in \cite{ArkaniHamed:2007ky} that the bound (\ref{vague_limit})  indeed holds
 for the classical inflaton trajectory in any theory of inflation that does not allow violations of the null energy condition,
 \[
 \rho+p\geq 0\;,
 \]
 where $\rho$ and $p$ are the energy density and the pressure. On the other hand, violation of the bound 
 (\ref{vague_limit}) is possible in theories able to violate the null-energy
 condition, such as ghost inflation \cite{ArkaniHamed:2003uz}. 
 This is actually encouraging and supports  arguments establishing the link
 between horizon complementarity and duration of inflation, as also the conventional black hole thermodynamics  breaks down in theories
 where the null energy condition can be violated \cite{Dubovsky:2006vk,Eling:2007qd}. 
 
There are two reasons leading to an uncertainty in the numerical value of the coefficient $c$ in the bound (\ref{vague_limit}) as proven in \cite{ArkaniHamed:2007ky}. First, at the time when the bound was proposed the exact condition for inflation to become eternal was unknown, 
and it was not even clear
whether there is a sharp distinction between eternal and non-eternal regimes. 
This issue was addressed in \cite{Creminelli:2008es}
 and the conclusion is that there is a sharp transition between these two regimes,
 with the condition not to have eternal inflation being
 \be
 \label{no_eternal}
 \Omega\equiv {2\pi^2\over 3}{\dot{\phi}^2\over H^4}\geq 1\;,
 \ee
 where $\dot\phi$ is the classical velocity of the inflaton field. 
 
 Now it is straightforward to find a bound on the number of $e$-foldings for the classical inflaton trajectory in single-field slow-roll
 inflation  in the non-eternal regime. Namely, one writes
 \be
 \label{precise_classical}
 {dS_{dS}\over dN_c}\equiv {M_{\rm Pl}^2dH^{-2}\over Hdt}=-{2M_{\rm Pl}^2\dot{H}\over H^4}=12\Omega\;,
 \ee
 where at the last step we made use of the second Friedmann equation. By integrating (\ref{precise_classical}) and using
 the condition (\ref{no_eternal}) for the absence of eternal inflation we obtain
 (\ref{vague_limit}) with the value $c=1/4$,
  \be
 \label{vague_limit1/4}
 3N_c\leq {S_{dS}\over 4}\;,
 \ee
 where $N_c$ is the number of $e$-foldings on the classical inflaton trajectory.

 However, this does not establish the sharp version of the bound (\ref{vague_limit}) yet. 
 There is  another reason for the uncertainty in the bound (\ref{vague_limit}). Namely,  the above analysis  (and that of  \cite{ArkaniHamed:2007ky}) is restricted to the
 classical inflaton trajectory. This approximation clearly breaks down when $\Omega$ is of order (but still larger than) one, so that inflation is close to be 
 eternal. In this case, even though inflation is not eternal, the typical inflaton trajectory is very different from the classical one and
 can be much longer. More generally, at the quantum level for any value of $\Omega$ there is always a non-vanishing probability for the
actual  inflaton trajectory to be long enough to violate the bound (\ref{vague_limit}) for any value of $c$.

In this situation it is natural to study what is the probability distribution for inflaton trajectories of different lengths---in other words, what is the probability distribution for the volume of the Universe $\rho(V)$ after inflation. 
It is not clear a priori what the natural generalization of the bound (\ref{vague_limit}) should look like at the quantum level.
One might expect that there exists  a value of $c$ such that the probability to violate (\ref{vague_limit}) is suppressed,
for example as non-perturbative quantum gravity effects $e^{-S_{dS}}$ (which would correspond to $\rho(V)\sim 1/V^\alpha$)
or even more, for example exponentially with the volume $\rho(V)\sim e^{-V}$ 
(which would correspond to order $e^{-e^{S_{dS}}}$ effects).
What we find from our analysis is that such value of $c$ exists (it is $c=1/2$) and that the probability associated to the violation
of the bound is actually super-exponentially small, \emph{i.e.} $\sim e^{-e^{ S_{dS}}}$.
%
%

To achieve this goal we obtained another result of independent interest. Namely, we calculated in an explicit form the probability distribution for the volume of the Universe after slow-roll inflation $\rho(V)$ both in eternal and
non-eternal regimes. This offers further insight in the actual geometry of the eternally inflating spacetime. While, unlike the density perturbation spectrum, this quantity is not of much interest for current observations, it still appears to be one of the natural ``theoretical" observables to look at in the study of eternally inflating Universes. 
In particular, according to \cite{Creminelli:2008es},
the order parameter for the transition to eternal inflation is the normalization of $\rho(V)$,
\be
P_{\rm ext}=\int_0^\infty dV\rho(V)\;.
\ee
At $\Omega>1$ this quantity is equal to 1, in agreement with the naive expectation. However, at $\Omega<1$ the normalization $P_{\rm ext}$ becomes smaller then 1, indicating that there is a non-zero probability $(1-P_{\rm ext})$ for the reheating volume to be infinite, {\it i.e.}
for inflation to last forever. In this paper we will rederive this result in yet another, somewhat more explicit, way.
It is also worth noting that recently the far exponential tail of the probability $\rho(V)$ was calculated in \cite{Winitzki:2008ph}
in the eternal regime. This result was used there to define a ``reheating-volume" measure for observables after eternal inflation.
It appears that the explicit expression for $\rho(V)$ has good chances to be useful in further theoretical studies of de~Sitter space and eternal inflation.

The rest of the paper is organized as follows. We start section~\ref{sec:bact2infl} with a review of the discrete stochastic
branching process introduced in  \cite{Creminelli:2008es} to describe inflation. Then we use this model
to derive a differential equation (\ref{eq:phi}) for the Laplace transform of the probability distribution $\rho(V)$.
 Similar discrete models were used in \cite{Winitzki:2005ya} and they are essentially equivalent to the stochastic description of inflation by
Starobinsky \cite{starobinsky}. Within the stochastic  approach equation (\ref{eq:phi}) is known as the ``non-linear Fokker--Planck 
equation" \cite{Winitzki:2001np}.

In section~\ref{sec:solutions} we provide  approximate solutions for the equation (\ref{eq:phi}) and calculate 
the probability distribution in different regimes. We start with a discussion of the general properties
of the solutions of (\ref{eq:phi}), and rederive in a new and very explicit way that $\Omega=1$ is the transition point to
the eternally inflating regime, {\it i.e} that $P_{\rm ext}=1$ for $\Omega>1$ and $P_{\rm ext}<1$ for $\Omega<1$.
Then in section~\ref{sec:moments}  we study the moments of the volume distribution. We show that there is a simple way of calculating them 
without actually solving  equation (\ref{eq:phi}) and performing the Laplace transform. In agreement with the results
of \cite{Creminelli:2008es} we prove that at $\Omega>1$  sufficiently high moments diverge for any value of $\Omega$
if the inflaton field is allowed to take arbitrarily high values. 
We find the values  $\Omega_n$ such that  the $n$-th moment diverges at $\Omega<\Omega_n$, and illustrate
our method by explicitly calculating the average and the variance.

In section~\ref{sec:classical} we start analyzing the properties of the probability distribution by performing the
Laplace transform.
First, we consider the semiclassical limit $\Omega\gg1$, and 
 find an approximate solution for eq.~(\ref{eq:phi}) in this limit. It turns out that because of the non-commutativity of
 the large-volume and the large-$\Omega$ limits this solution does not capture correctly the large volume tail of the probability distribution where the probability becomes smaller than $\sim e^{-\Omega}$.
 The study of this solution is also instructive for developing an intuition on how to perform the
 Laplace transform of the solutions of eq.~(\ref{eq:phi}).
%
In section~\ref{sec:Wgen} we apply this 
intuition for general $\Omega>1$, when  it is not possible to find an approximate solution to (\ref{eq:phi}) in closed form.
By solving this equation in different regimes one can obtain enough information to reconstruct
 the probability
distribution in the physically relevant case $N\gg1$.
$\rho(V)$ turns out to be peaked around the average value
 \begin{eqnarray}
 \label{Nav}
  {\overline N}={2N_c\over 1+\sqrt{1-\Omega^{-1}}}\;. 
 \end{eqnarray}
For $N<{\overline N}$ it takes the following Gaussian form
\be
\label{gaussian}
\rho(N)\simeq {\cal N}e^{-{(3N-3{\overline N})^2\over 2\sigma^2}}\;,\qquad \Omega > 1
\ee
 where  the width $\sigma$ is equal to
 \begin{eqnarray}
 \sigma^2={2\over\Omega(1+\sqrt{1-\Omega^{-1}})^2}\;.
 \end{eqnarray}
As $\Omega$ approaches the transition point, $\Omega=1$, the width $\sigma$ becomes of order one, 
which is still narrow in the regime of large number of $e$-foldings, $N\gg 1$, the one we are interested in. 
So the most important consequence of the change of $\Omega$ is that the average number of $e$-foldings ${\overline N}$ changes.
In agreement with the naive expectation it increases as $\Omega$ approaches the transition point, but
it does not grow a lot: at $\Omega=1$  the average number of $e$-foldings is twice as large as the classical one.

At large volumes, $N\gtrsim {\overline N}$, the probability distribution becomes exponential in $N$ (or, equivalently,
power-law in the volume $V$),
\be
\label{exp}
\rho(N)\propto e^{-6\Omega N\l(1+\sqrt{1-{1\over \Omega}}\r)}=V^{-2\Omega \l(1+\sqrt{1-{1\over \Omega}}\r)}\;.
\ee

\begin{figure}[t]
\begin{center}
\includegraphics[width=14cm]{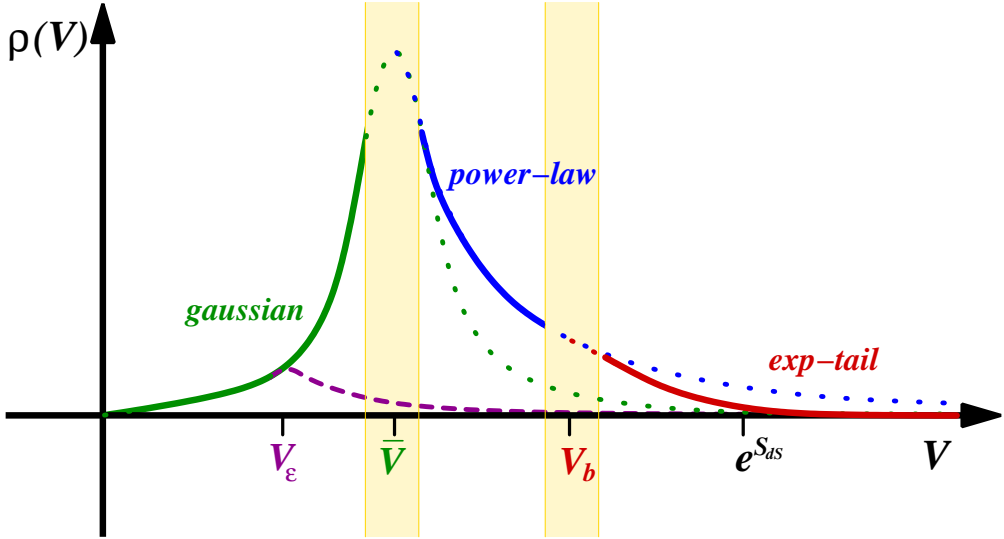} 
\caption{\label{fig:rhoV-proto} \small \it Typical shape for the probability distribution of the volume $\rho(V)$.
For small volumes the behavior is gaussian with the number of $e$-foldings ($\rho\sim e^{-c(N-\overline N)^2}$); for volumes larger
than the average value $\overline V$, $\rho(V)$ follows a power law in the volume ($\rho\sim 1/V^\alpha$) that eventually
turns into an exponential law ($\rho \sim e^{-{\rm const}\cdot V}$) at large enough volumes ($V\gtrsim V_b$).
When $\Omega<1$ the exponential tail starts earlier at $V\simeq V_\epsilon=e^{\pi/(2\sqrt{1-\Omega})}$.}
\end{center}
\end{figure}

Then we proceed with the eternal inflation regime. First, we study what happens
in the vicinity of the transition point, {\it i.e.} when $\Omega=1-\epsilon$ with $0<\epsilon\ll 1$.
We find that the probability distribution $\rho(N)$ 
is not changed until $N\sim\pi/(6\sqrt{\epsilon})$.
At large volumes, $N\gtrsim\pi/(6\sqrt{\epsilon})$,
 it becomes much more strongly suppressed, $\rho(N)\propto e^{-{\rm const} \cdot e^{3N}}$. 
 This behavior is easier to interprete  in terms of the volume distribution $\rho(V)$ (rather
than the $e$-folding distribution $\rho(N)$).
It  indicates that if the volume gets large enough, $V\gtrsim e^{\pi/(2\sqrt{\epsilon})}$, the probability for inflation
to terminate is exponentially small $P_{\rm ext}\propto e^{-{\rm const}\cdot V}$.
Related to that, we find that the total probability for inflation to terminate is of order $e^{-\Omega(3N_c)^2}$
(this also applies for $\epsilon\sim{\cal O}(1)$), indicating that it is saturated by the small-$N$ tail of the Gaussian distribution (\ref{gaussian}). 
This behavior smoothly matches with yet another regime where one can find
the probability distribution explicitly---$\Omega\simeq 0$. Here the probability distribution $\rho(V)$ is exponentially small $\rho(V)\propto e^{-V/2}$ for all volumes of interest, $V\gg 1$.

All these considerations were made in the approximation where the inflaton potential goes up to arbitrary high
values of the inflaton field. This is clearly unrealistic and we 
 conclude in section~\ref{sec:solutions} by discussing what happens in the presence of a ``barrier" at large values of the inflaton field. As expected, the presence of a barrier affects only the far tail of the volume distribution by making it exponentially suppressed at large volumes, $\rho(V) \propto e^{-{\rm const}\cdot V}$, for any value of $\Omega$. 
 Note that if the initial value of the inflaton field is not too close to the barrier, this effect is relevant only for
 inflaton trajectories whose probability is smaller than the uncertainty coming from non-perturbative
quantum gravity effects ($\sim e^{-S_{\rm dS}}$).
The various behaviors of the probability distribution are summarized in fig.~\ref{fig:rhoV-proto}
that shows the shape of $\rho(V)$ in the different regimes of the volume.

These results imply that the quantum version of the bound (\ref{vague_limit}) does hold with $c=1/2$.
In the concluding section~\ref{conclusions} we give the physical explanation of the behavior of the probability distribution
and show that an inflaton trajectory  with more than $S_{dS}/6$ 
$e$-foldings and such that inflation terminates globally in the entire space is
super-exponentially improbable.
We also speculate on the possibility that the value $c=1/2$ for the coefficient
in (\ref{vague_limit}) that we obtained in our analysis might have a natural physical interpretation. 
%
In the appendix we cross check our results by calculating the average volume directly from the inflaton stochastic equations.

\section{From bacteria to inflation} \label{sec:bact2infl}
As explained in the introduction, we want to calculate and study the probability distribution $\rho(V,\phi)$ 
of the reheating volume given a certain initial value of the field $\phi$. 
This calculation does not seem to be straightforward, as the only available definition of the distribution is a 
rather formal functional integral formula \cite{Creminelli:2008es} 
\be
\label{fi}
\rho(V,\phi)=\int \! {\cal D}\bar\phi \: {\cal P}[\bar\phi,\phi] \: \delta \l[ V- \int \!d^3x \, e^{3Ht_r
(\vec x)} \r] \;,
\ee
where ${\cal D}\bar\phi$ is some vaguely defined measure on the set of all possible space-time realizations
of the inflaton field, ${\cal P}[\bar\phi,\phi]$ is the probability of a specific realization and $t_r(\vec x)$ is 
the reheating time for a given realization as a function of the comoving coordinate 
$\vec x$.  Evaluating directly the functional integral is of course a very hard task. 
As usual with functional integrals, 
in order to gain more control it is natural to switch to a discretized description of the 
inflationary dynamics. This approach has been recently developed in \cite{Creminelli:2008es} and
similar models have also been studied in the context of eternal inflation 
in \cite{Winitzki:2005ya}.
Up to small extensions, 
section~\ref{sec:review} is mainly 
a review of the results in \cite{Creminelli:2008es}, which we use
to derive the formula for the probability distribution $\rho(V,\phi)$ in section~\ref{sec:volume equation}. 
The resulting solution for $\rho(V,\phi)$ will be discussed in section~\ref{sec:solutions}.
As a cross-check of the method, in the appendix we also present a direct computation of the volume average.

\subsection{\label{sec:review}Review of bacteria model}

\begin{figure}[ht!!]
\begin{center}
\includegraphics[width=12cm]{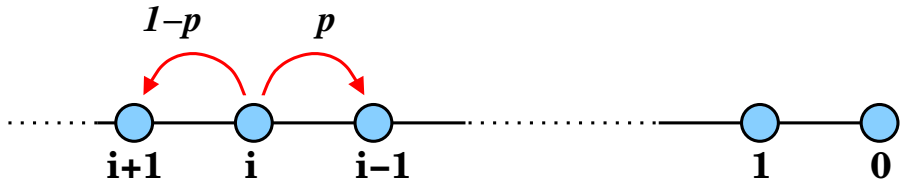}
\caption{\label{fig:branching process} \small \it The branching process.}
\end{center}
\end{figure}

With a biological analogy, consider at $t=0$ a bacterium that can live in a discrete set of positions along 
a line (see fig.~\ref{fig:branching process}). At $t=1$ the bacterium replicates into $N_r$ copies. Then, each 
bacterium (independently of all the others) hops with probability $p$ to the neighboring site on its right, 
and with probability $(1-p)$ on the left.  $N_r$ and $p$ are fixed numbers. At $t=2$ each second-generation 
bacterium reproduces itself, and so on. The analogy with the inflationary system is clear: 
each bacterium represents an Hubble patch; sites are inflaton values. Reproduction is the analogue of 
the Hubble expansion; at  every $e$-folding $\sim e^3$ new Hubble volumes are produced starting from 
one. From then on the inflaton inside each Hubble volume evolves independently, with a combination of 
classical rolling and quantum diffusion. This is represented by
the random hopping of our bacteria. The difference in the probabilities of moving right and left gives a 
net drift, and thus corresponds to the classical motion. To complete the analogy we have to assume that 
there is a ``reheating'' site, $i=0$ in the figure: when a bacterium ends up there it stops to reproduce and 
to move around---it dies.
In the bacteriological analogy the reheating volume corresponds to the number of dead bacteria 
(= non-reproducing Hubble patches) in the asymptotic future. For analogy we denote the latter quantity by $V$, 
which of course now takes discrete values. Our task is to study the probability distribution of $V$ as a 
function of the parameter $p$.
A discrete system like the one we described goes under name of {\em branching process}, more 
precisely a multi-type Galton-Watson process (see e.g.~ref.~\cite{branching_books}).

To connect to the inflationary case, if $\phi$ is the inflaton, and $t$ is the time of the FRW metric, one can 
make the following identifications in terms of the position $j$, time-step $n$, field-space interval $\Delta 
\phi$ and time interval $\Delta t$,
\begin{eqnarray} \label{eq:jndef}
j=\frac{\phi}{\Delta\phi}\ , \quad \quad \quad n=\frac{t}{\Delta t} \;.
\end{eqnarray}
By taking the continuum limit $\Delta \phi, \Delta t \rightarrow 0$ in such a way that 
\be
\label{dtdf_relation}
\Delta t={4\pi^2\over H^3} (\Delta\phi)^2\ ,
\ee
where $H$ is the Hubble rate in the inflationary process, and by defining $N_r$ as 
\be \label{eq:Nrlim}
N_r=1+3 H\Delta t \ ,
\ee
and $p$ by the relationship 
\be
\label{p_matching}
(1-2p)\frac{\Delta \phi}{\Delta t}=\dot \phi\; \quad \Rightarrow \quad p={1\over 2}+\sqrt{6\pi^2\Omega} \, {\Delta\phi\over H}\ ,
\ee
the equation for the probability of a bacterium to be at site 
$j$ at time $n$ becomes the stochastic equation for the inflaton \cite{starobinsky,Linde:1986fd,Linde:1986fc}
\be
\frac{4\pi^2}{H^3} \de_{t} P(\bar\phi,t)=\frac12 \de^2_{\bar\phi} P(\bar\phi,t)+\frac{2\sqrt{6\pi^2\Omega}}{H} \de_{\bar\phi} P(\bar\phi,r)\,,
\ee
where $P(\bar\phi,t)$ is the probability that the inflaton $\phi$ after a time $t$ has a value $\bar \phi$.
In \cite{Creminelli:2008es} general arguments
establishing the matching with the continuum limit were presented and checked in several calculations performed 
both in terms of the bacteria model and of the inflaton.

Let us therefore study in more detail the bacteria model, and consider a branching process on a line of 
length $L$. 
 A convenient tool to study the branching process is the set of generating functions $f^{(n)}_i (s_j)$, 
where
 $i,j=0,\dots, L$. These are defined as power series
 \be
 \label{fs_def}
 f^{(n)}_i(s_j)=\sum_{k_1\dots k_L} p^{(n)}_{i; k_0\dots k_L}s_0^{k_0}\dots s_L^{k_L}\,,
 \ee
where $p^{(n)}_{i; k_0\dots k_L}$ is the probability that, in a branching process that started with a single
bacterium at the $i$-th site after $n$ steps, one has $k_0$ bacteria at the zeroth site, $k_1$ bacteria
at the first site, etc.  It is convenient to combine together 
 all functions $f^{(n)}_i$ with the same number
of steps $n$ into a map $F_n$ from the $L+1$-dimensional space of the auxiliary parameters $s_i$
into an $L+1$-dimensional space parameterized by the $f_i$'s. Also in what follows we will sometimes 
drop
the subscript from the $s_i$ variables and denote by $s$ a point in the $L+1$-dimensional space
with coordinates $(s_0,\dots, s_L)$.
For example, for a branching process of the sort as described in fig.~(\ref{fig:branching process}), $F_1
$ is given by
\begin{eqnarray}\label{eq:f^1}
f_0^{(1)}(s_0,\ldots,s_L)&=&s_0\,,\\ \nonumber
f_1^{(1)}(s_0,\ldots,s_L)&=&\left((1-p)s_2+p\, s_0\right)^{N_r}\,,\\ \nonumber
&\vdots& \\ \nonumber
f_{i}^{(1)}(s_0,\ldots,s_L)&=&\left((1-p)s_{i+1}+p\,s_{i-1}\right)^{N_r}\,,\\ \nonumber
&\vdots& \\ \nonumber
f_{L}^{(1)}(s_0,\ldots,s_L)&=&\left((1-p) s_{L}+p\, s_{L-1}\right)^{N_r}\,, \nonumber 
\end{eqnarray}
where we have made a specific choice of boundary conditions at the site $L$, which 
we will refer to as the ``barrier''. The particular choice of the boundary condition will affect only 
very marginally our results.
The main property making generating functions useful is the iterative relation
\be
\label{iterative}
F_{n+1}=F_1(F_n)\;.
\ee
This property is straightforward to check by making use of the definition of the
branching process and elementary 
properties of probabilities.

We will be interested in the late time behavior of the branching process, which is determined
by the asymptotic function $F_{\infty}$. The iterative property (\ref{iterative}) implies that 
\begin{equation}\label{eq: fixed point}
F_1(F_\infty)=F_\infty\;,
\end{equation}
{\it i.e.}~the set of values of the function $F_\infty$ is a subset of the fixed points
of the function $F_1$, such that
\be
\label{fixed}
F_1(s)=s\;.
\ee
For our purposes it is enough to study the mapping $F_1$ inside the $L+1$-dimensional cube $I_{L+1}$ 
of unit size,
$0\leq s_i<1$. The definition (\ref{fs_def}) of the generating functions implies that all partial derivatives of
$F_1(s)$ are positive. Also, the normalization of probabilities implies that
\[
F_1(1,\dots,1)=(1,\dots, 1)\equiv \vec{1}\;.
\]

\begin{figure}[t!]
\begin{center}
\includegraphics[width=13cm]{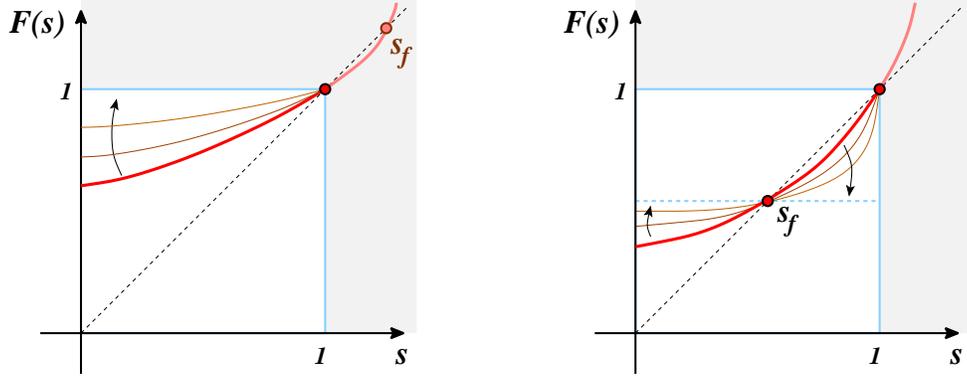}
\caption{\label{extinction} \small \it {\em Left:} Plot of $F_1(s)$ restricted to the hypercube $I_L$ and with 
$s_0=1$, for large $p$ {\em (thick curve)}. The only fixed point  in the unit cube is $s=1$. Further 
applications of $F_1$ {\em (thinner curves)} drive the curve to the $F_{\infty} = 1$ line. {\em Right}: For 
smaller $p$'s a new fixed point $s_f$ enters the unit cube.  Now the limiting line is $F_{\infty} = s_f$.}
\end{center}
\end{figure}

In the bacteria model, it is rather straightforward to see how the transition to the eternal inflationary 
regime happens. For this, it is useful to restrict the function $F$ to the $L$-dimensional hypercube $I_L$ 
with $s_0=1$, which amounts to marginalizing over the number of dead bacteria (see eq.~(\ref
{fs_def})). Note that, if 
the mapping $F_1$ has no other fixed points in the cube $I_L$ apart from $\vec{1}$ (see fig.~\ref
{extinction}), then 
\begin{equation}\nonumber
\left. F_\infty\right|_{s_0=1}=\vec{1}\;.
\end{equation}
By definition of the generating functions, eq.~({\ref{fs_def}}), this means that in the late time asymptotics 
with
probability one there are no alive bacteria at any of the sites. The extinction probability is exactly equal 
to
one (inflation ends). The situation changes when  a non-trivial fixed point $s_f$ solving eq.~(\ref{fixed}) 
enters the region $I_L$ (see fig.~\ref{extinction}). 
Now one has 
\be
\label{1sf}
\left. F_\infty\right|_{s_0=1}=(1,s_f)<\vec{1}\;.
\ee
This implies that, as before, the probability to have any finite non-zero  number of alive bacteria at any 
site vanishes. However, the probabilities to have zero bacteria at the various sites, 
\be
p^{(\infty)}_{i;{{\rm any}},0\dots0} = f_i^{(\infty)}(0)=(s_f)_i \,, \quad \quad i\neq 0\ ,
\ee
are all less than one.  This means that there is a non-vanishing probability that the population never dies 
out  and that the total number of bacteria grows indefinitely
at late times. This corresponds to the eternal inflation regime. Clearly, this implies that the number of 
dead bacteria also has a finite probability to grow indefinitely; in the context of eternal inflation this 
translates into
\[
\int_0^\infty dV \rho(V,\phi) < 1\,.
\]
In the continuum limit and with infinite barrier this transition happens at $\Omega=1$.

We would like to stress that $F_\infty$ is a function of  $s_0$ only. 
Mathematically this follows from the convexity of $F_1$ with respect to $s_i$, $i\neq0$, from the fact that 
$F_1$ is defined between 0 and 1, and that $F_1(\vec{1})=\vec 1$  (see fig.~\ref{extinction}). This can 
also  be intuitively understood in the following way: in the non-eternal regime, in the infinite future there 
is zero probability of having any bacteria alive, so $F_\infty$ can not depend on any of the $s_i$, $i\neq0
$. In the eternal inflation case, the population either extinguishes or becomes infinite. In either case, the 
probability of finding a finite number of bacteria at any site is zero, which again forbids any dependence 
on $s_i$, $i\neq0$. Therefore, $F_\infty=F_\infty(s_0)$. 
Eq. (\ref{eq: fixed point}) for the fixed point then becomes
\begin{eqnarray}
f_0^{(\infty)}(s_0)&=&s_0\,,\nonumber \\ \nonumber
&\vdots& \\ \nonumber
f_{i}^{(\infty)}(s_0)&=&\left((1-p)f_{i+1}^{(\infty)}(s_0)+p\,f_{i-1}^{(\infty)}(s_0)\right)^{N_r}\,,\\ \nonumber
&\vdots& \\ 
f_{L}^{(\infty)}(s_0)&=&\left((1-p) f_{L}^{(\infty)}(s_0)+p\, f_{L-1}^{(\infty)}(s_0)\right)^{N_r}\,. \label{eq:discr-syst}
\end{eqnarray}
This is the set of equations that determine $f_i^{(\infty)}(s_0)$. Once $f_i^{(\infty)}(s_0)$ is found, 
one can extract the asymptotic probability distribution $p^{(\infty)}_{i;k_0}$ using eq.~(\ref{fs_def}).

\subsubsection{\label{sec:2-site}An example: the 2-sites case}
\begin{figure}[t!]
\begin{center}
\includegraphics[width=3.4cm]{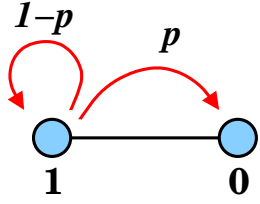}
\caption{\label{fig:2_site_line} \it \small The 2-site branching process.}
\end{center}
\end{figure}

We find it instructive to illustrate the general formalism using a simple explicit example.
Consider the minimal branching process with just two sites and ${N_r}=2$ copies at each 
reproduction event (see fig.~\ref{fig:2_site_line}).
In this case the generating functions (\ref{fs_def}) are particularly simple,
\begin{eqnarray}
&&f^{(1)}_0 (s_0,s_1)=s_0 \ , \nn \\
&&f^{(1)}_1 (s_0,s_1)=\big((1-p)s_1+p s_0 \big)^2 \ . \label{eq:map}
\end{eqnarray}
It is straightforward to apply here the generic results discussed in the last section. $f^{(\infty)}_i(s_0)$ is 
given by the fixed point of the mapping defined in eq.~(\ref{eq:map}) inside the unit interval $0 \le s_i \le 1$. We obtain
\begin{eqnarray}
&&f^{(\infty)}_0 (s_0)=s_0 \ , \\
&&f^{(\infty)}_1 (s_0)=\frac{ 1-2p s_0+ 2p^2 s_0- \sqrt{1-4p(1-p)s_0} }{ 2(1-p)^2 }\ .
\end{eqnarray}
The extinction probability is given by  
\be
P_{\rm ext} = \sum_{k=0}^{\infty} p_k= f^{(\infty)} _1(1)
= \frac{ 1-2p + 2p^2 - \sqrt{(1-2p)^2} }{ 2(1-p)^2 } =
\left\{ \begin{array}{ccc}
	1 & & p > 1/2 \\
	(\frac{p}{1-p})^2 & & p < 1/2
	\end{array} \right. \; ,
\ee
where $p_k\equiv p^{(\infty)}_{1;k\,0}$. The extinction probability indeed drops below one for $p<1/2$. 
We get the quite intuitive result that the critical probability is
$p_c=1/2$.
By Taylor expanding $f^{(\infty)}_i(s_0)$, we can obtain the probability of the volume $p_k$ as in eq.~(\ref{fs_def})
\begin{equation}
f_1^{(\infty)}(s_0)=\frac{1}{2(1-p)^2}\sum_{k=2}^{\infty}\frac{(2k-3)!}{2^{2k-2}k!(k-2)!}(4p(1-p)s_0)^k \ ,
\end{equation}
which allows us to extract the large-$k$ asymptotic of the probability distribution $p_k$,
\begin{equation}\label{eq:2-site-prob}
p_k\sim \frac{1}{(1-p)^2}\ \frac{1}{k^{3/2}}e^{-k\log(\frac{1}{4p(1-p)})}\,.
\end{equation}
We see that the probability for large volumes goes to zero exponentially fast for any $p\neq p_c=1/2$ 
(notice that $4p(1-p)\leq1$ for any $p$, and becomes equal to one only for $p=1/2$). At $p=p_c$ the 
exponential suppression disappears and we are left with a power law behavior, 
corresponding to the singularity of $f_1^{(\infty)}(s_0)$ at $s_0=1$, as a result the probability distribution becomes not normalized
for $p\leq p_c$. This can be 
understood in the following way. So far we have studied the probability distribution in the infinite time limit.
If instead we let the process go on only for a finite number of time steps $n$, the 
probability distribution develops a bump at large volume (see \cite{Creminelli:2008es} for 
details), which can be roughly thought of as a piece proportional to $\Theta(1-P_{\rm ext}) \delta(k-2^{n})$. 
For $p\leq p_c$, in the limit $n\to\infty$, the $\delta-$function migrates to infinity and 
makes all the moments of the volume diverge. In our calculation, we can not see such a $\delta$-function because we have already taken the limit of infinite time steps, but still we are left with a non-normalized distribution. Because of this
even for $p<p_c$ all the moments of the volume diverge even 
if the probability distribution goes exponentially fast to zero at infinity,.

\subsection{\label{sec:volume equation}The equation for $\rho(V)$}
We want now to take the continuum limit of the bacteria model as explained in the beginning of section~\ref{sec:review}.
According to eq.~(\ref{eq:jndef}), we define $f^{(\infty)}(\phi;s_0)=f_{\phi/\Delta\phi}^{(\infty)}(s_0)$ so that the 
system in eq.~(\ref{eq:discr-syst}) now becomes
\begin{eqnarray}
f^{(\infty)}(0,s_0)&=&s_0\,,\\ \nonumber
&\vdots& \\ \nonumber
f^{(\infty)}(\phi;s_0)&=&\left((1-p)f^{(\infty)}(\phi+\Delta\phi;s_0)+p\,f^{(\infty)}(\phi-\Delta\phi;s_0)\right)^{N_r}\,,\\ \nonumber
&\vdots& \\ \nonumber
f^{(\infty)}(\phi_b;s_0)&=&\left((1-p) f^{(\infty)}(\phi_b;s_0)+p\, f^{(\infty)}(\phi_b-\Delta \phi;s_0)\right)^{N_r} \,,
\nonumber
\end{eqnarray}
where $\phi_b$ represents the value of $\phi$ at the barrier.

By taking the limit $\Delta\phi\to 0$ $\Delta t\to 0$ according to the prescription given in eqs.~(\ref{dtdf_relation}), 
(\ref{eq:Nrlim}), and (\ref{p_matching}) we obtain the following second order differential equation
\begin{equation}\label{eq:phi}
\frac{1}{2}\frac{\partial^2}{\partial \phi^2} f^{(\infty)}(\phi;s_0) -\frac{2\pi\sqrt{6\Omega}}{H}\frac{\partial}
{\partial \phi} f^{(\infty)}(\phi;s_0)+\frac{12\pi^2}{H^2} f^{(\infty)}(\phi;s_0)\log \left[ f^{(\infty)}(\phi;s_0)
\right]=0\ ,
\end{equation}
with the following two boundary conditions
\begin{eqnarray}
f^{(\infty)}(0;s_0)&=&s_0 \ ,\\ \nonumber
\left.\frac{\partial}{\partial \phi}f^{(\infty)}(\phi;s_0)\right|_{\phi_b}&=&0 \,.
\end{eqnarray}
in agreement with \cite{Winitzki:2001np,Winitzki:2008ph} where the same 
equation has been obtained in a different way.
The second derivative term in eq.~(\ref{eq:phi}) comes from the quantum fluctuations of the inflaton 
(equivalent to the random jumps of the bacteria), the first derivative term comes from the classical drift (in fact it is 
proportional to $\Omega$), and the $\log$ in the last term comes from the production of independent 
Hubble patches (equivalent to the reproduction of bacteria).

This equation will be central in the rest of paper:  $f(\phi;s_0)$ is the Laplace transform of the probability 
of obtaining a certain reheating volume starting from any initial value of $\phi$. Indeed, in the 
discrete model, from the definition in eq.~(\ref{fs_def}), the generating function at infinite time is 
connected to the probability $p_{j,k}$ of having 
$k$ dead bacteria at infinite time starting with one bacterium at the site $j$ by
\be\label{eq:finftysum}
f_j^{(\infty)}(s_0)=\sum_{k=0}^\infty p_{j,k}\, s_0^k\,.
\ee
Taking the continuum limit we get
\be \label{eq:rhotof}
f^{(\infty)}(\phi;s_0)=\int_0^\infty dV \rho(\phi,V)\, s_0^{V}\,.
\ee
This expression can be inverted to obtain the probability distribution for the volume
\be\label{eq:Laplace}
\rho(\phi,V)=\frac{1}{2\pi i}\int_{\gamma-i\infty}^{\gamma+i\infty} d\left(-\log(s_0)\right) f^{(\infty)}(\phi;s_0)e^{- V\log
(s_0)}
\ee
where $\gamma$ must be chosen such that Re${(\gamma)}>-\log(s^{\rm sing}_0)$ for any singularity 
$s^{\rm sing}_0$ of $f(\phi;s_0)$~\footnote{
From eq.~(\ref{eq:rhotof}) we see that $f^{(\infty)}(\phi;s_0)$ cannot have singularities for Re$(\log(s_0))<0$,
therefore eq.~(\ref{eq:Laplace}) holds for every $\gamma>0$.}. 
Notice that in eq.~(\ref{eq:Laplace}) we need to analytically continue the function $f^{(\infty)}(\phi;s_0)$ 
to unphysical values of $s_0$ ($s_0\notin[0,1]$), which actually dominate the integral at large volumes 
as we will see in section~\ref{sec:solutions}.
We have therefore 
obtained a procedure to compute the probability distribution of the reheating volume: solve the 
differential equation (\ref{eq:phi}), and then perform the anti-Laplace transform (\ref{eq:Laplace}).

%

\section{Probability distribution of the volume after inflation} \label{sec:solutions}
In the previous section we saw that the probability distribution of the volume can be calculated in two steps. The first is to solve the differential equation (\ref{eq:phi}). For convenience we rewrite it here as
\be \label{eq:diffeq}
{\ddot f}(\tau;z)-2\Oh {\dot f}(\tau;z)+f(\tau;z)\log [f(\tau;z)]=0\,,
\ee
where the dot represents a partial derivative with respect to $\tau$ and
$f(\tau;z)$, $\tau$ and $z$ are related to 
$f^{(\infty)}(\phi;s_0)$, $\phi$ and $s_0$ of the previous section via
\bea
f(\tau;z)&\equiv& f^{(\infty)}(\phi;s_0)\,, \\
\tau&\equiv& 2 \pi \sqrt6 \frac{\phi}{H}=6\Oh N_c\,, \\
z&\equiv& -\log(s_0)\,,
\eea
with $N_c$ being the classical number of $e$-foldings ($N_c\equiv -H \phi/\dot\phi$).
The solution $f(\tau,z)$ has also to satisfy the following boundary conditions
\be \label{eq:bc1}
f(0;z)=s_0=e^{-z}\,,
\ee
\be \label{eq:bc2}
\dot f(\tau_b;z)=0\,,
\ee
and the constraint
\be
f(\tau;z)\in [0,1]\,.
\ee

The second step  is to calculate the integral
\be \label{eq:rhoV}
\rho(V,\tau)=\frac{1}{2\pi i}\int_{0^+-i\infty}^{0^++i\infty} dz\, f(\tau;z)e^{z V}\,,
\ee
that gives the probability distribution $\rho(V,\tau)$ to find a volume $V$ at the end of an
inflationary phase that started with the inflaton at the position $\phi=H \tau/(2\pi\sqrt6)$.
Recall that the volume $V$ that enters in eq.~(\ref{eq:rhoV}) is dimensionless
because it has been rescaled by the initial volume $V_0$ before inflation, i.e. $V={\rm Vol}/V_0$, with
Vol being the physical volume.
Notice that some properties of $\rho(V)$ can be obtained without evaluating the integral (\ref{eq:rhoV}),
but just by studying the solution $f(\tau;z)$ around the point $z=0$. Indeed the momenta of the distribution
are simply related to the Taylor coefficients of $f(\tau;z)$ around $z=0$ since
\be \label{eq:momentsdef}
\la V^n \ra = \int_{0}^\infty dV\, V^n \rho(V,\tau)=(-1)^n\l.\frac{\de^{n}f(\tau;z)}{\de z^n}\r|_{z=0}\,.
\ee
It follows that the total probability to exit inflation globally is just fixed by $f(\tau;0)$,
\be \label{eq:PV}
P_{\rm ext}\equiv \int_{0}^\infty dV \rho(V,\tau)=f(\tau;0)\,.
\ee

\begin{figure}[t!]
\begin{center}
\includegraphics[height=7cm]{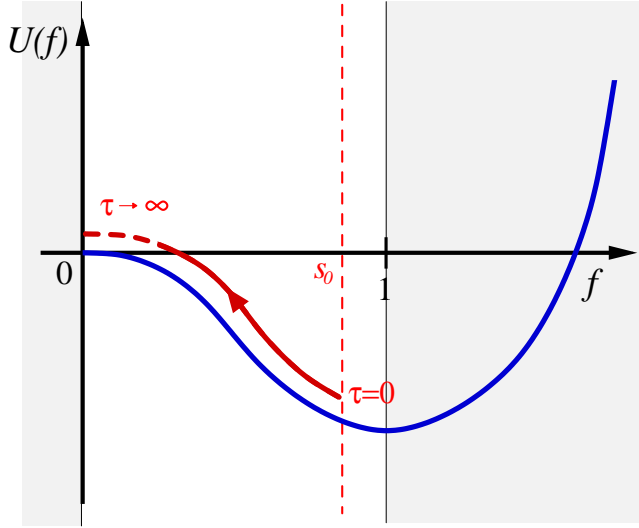}
\caption{\label{fig:pot} \small \it The potential $U(f)$ of eq.~(\ref{eq:Uf}). In the limit $\tau_b\to\infty$, 
the solution of the mechanical problem
(\ref{eq:diffeq}) represents the motion of a particle starting at $\tau=0$ in $f=s_0$, rolling uphill with an anti-friction
term and stopping on the top $f=0$ in an infinite time.}
\end{center}
\end{figure}

We proceed now with the study of the solutions of eq.~(\ref{eq:diffeq}).
This equation describes the motion
of the particle in a potential 
\be \label{eq:Uf}
U(f)=\frac{f^2}4\l(\log f^2-1\r)\,,
\ee
with an anti-friction term (see fig.~\ref{fig:pot}).  Unfortunately an explicit solution to eq.~(\ref{eq:diffeq}) is not known. 
However we will still be able to recover many properties of $\rho(V,\tau)$  by analyzing the analytic structure of the solution and by studying the problem in different limits.
The boundary conditions (\ref{eq:bc1}) and (\ref{eq:bc2}) as well as the condition $f \in [0,1]$
constrain the solution to start from the point $f=s_0$ at $\tau=0$
and to travel up-hill up to some point $f_{*}\in (0,s_0)$ where the velocity reaches zero at time $\tau_b$.  
We will focus on the case where $\tau_b\rightarrow+\infty$, and come back to the case of finite $\tau_b$ at the end of this section. In the $\tau_b\rightarrow+\infty$ case, the solution has to reach zero velocity only after an infinite time, and in doing so, it has to stay always within the interval $f \in [0,1]$. The only way to achieve this is for the solution to reach the top of the hill $f=0$ in infinite time. 

Let us see that such a solution exists for any $s_0<1$. Clearly for large enough initial velocities the solution overshoots the maximum and goes to the $f<0$ region, while for small enough velocities, the solution does not reach the top. 
Therefore, there is a critical initial velocity separating these two regimes
 such that the solution stops at $f=0$ in an infinite time.
    This can also be explicitly verified by finding, as we will show  later, an asymptotic form of this solution in the region $0<f\ll1$,
\be
f\sim e^{-\frac{1}{4}(\tau+\tau_0)^2}\ ,
\ee
where $\tau_0$ is an integration constant. 

The case $s_0=1$ is different.
For a solution that starts at $f=1$ at early times
we can approximate the potential $U(f)$ with an harmonic oscillator to obtain  
\be
\label{harmonic}
\ddot f-2\Oh \dot f + f-1=0\,,
\ee
whose general solution is of the form
\be \label{eq:flinintro}
f=1-e^{\Oh \tau}\l(A e^{\sqrt{\Omega-1}\tau}+Be^{-\sqrt{\Omega-1}\tau}\r)\,.
\ee
This shows that for $\Omega>1$ the solution is 'over-anti-dumped' and does not have a turning point, while for $\Omega<1$ the solution can oscillate. 
This behavior persists at the non-linear level as well---it is straightforward to check that the turning force due to the
potential (\ref{eq:Uf}) is always smaller than the turning force due to the harmonic potential in (\ref{harmonic}), so that
no turning point exists for $\Omega>1$ for the non-linear mechanical problem (\ref{eq:diffeq}) as well.
Consequently, at $s_0=1$ the solution that stops on the top of the hill exists only at $\Omega<1$. This
  solution describes  a non-trivial fixed point (\ref{1sf}). Its presence indicates that inflation is eternal at $\Omega<1$.
  Instead, at $\Omega>1$ the solution that reaches the top of the hill in an infinite time $\tau\to +\infty$
  starts at $f=1$ in the infinite past, $\tau=-\infty$; while all solutions that start at $f=1$ at finite time
  overshoot the top of the hill. 
  
\begin{figure}[t!]
\begin{center}
\includegraphics[height=7cm]{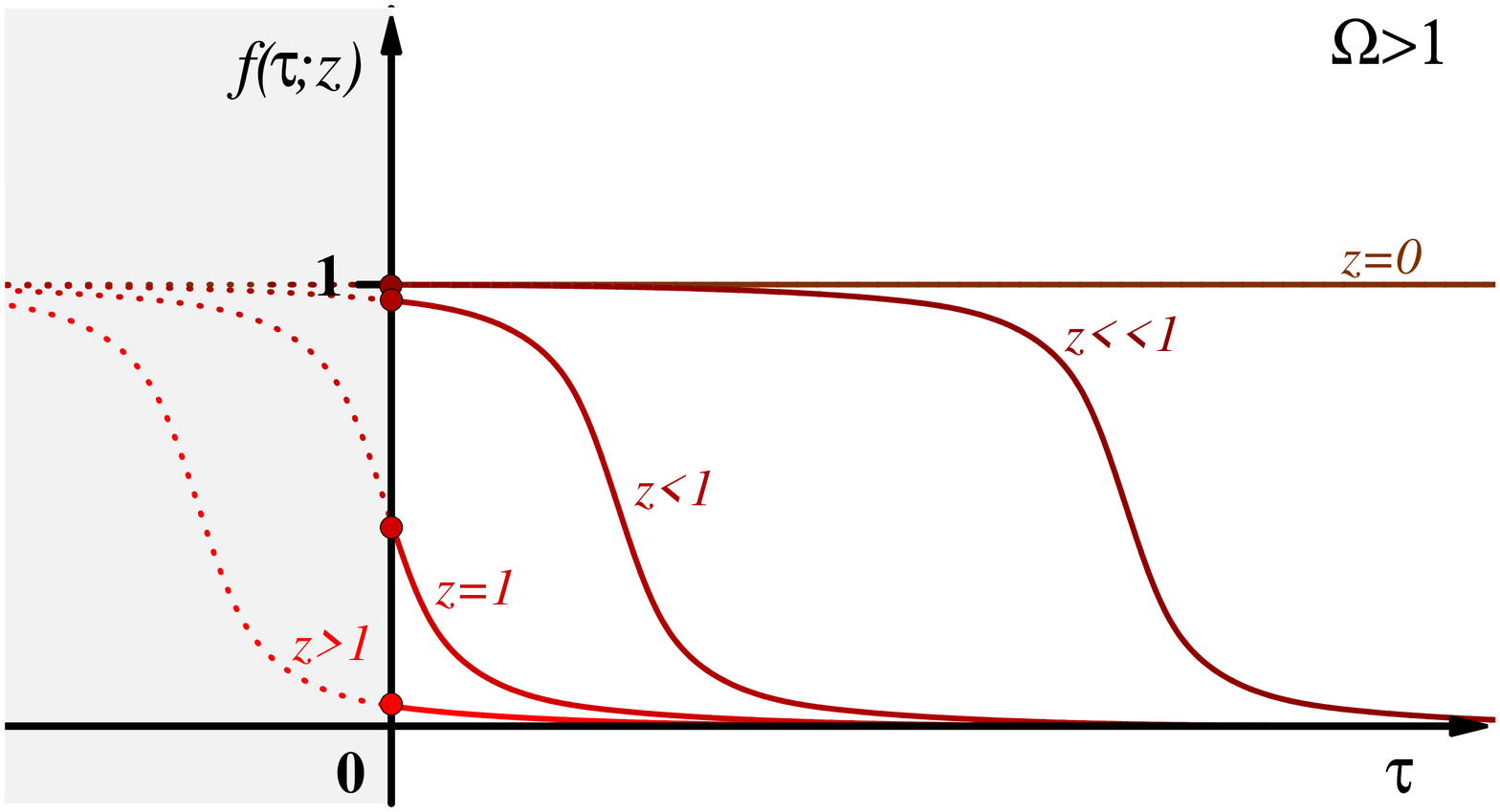}\\~\\
\includegraphics[height=7cm]{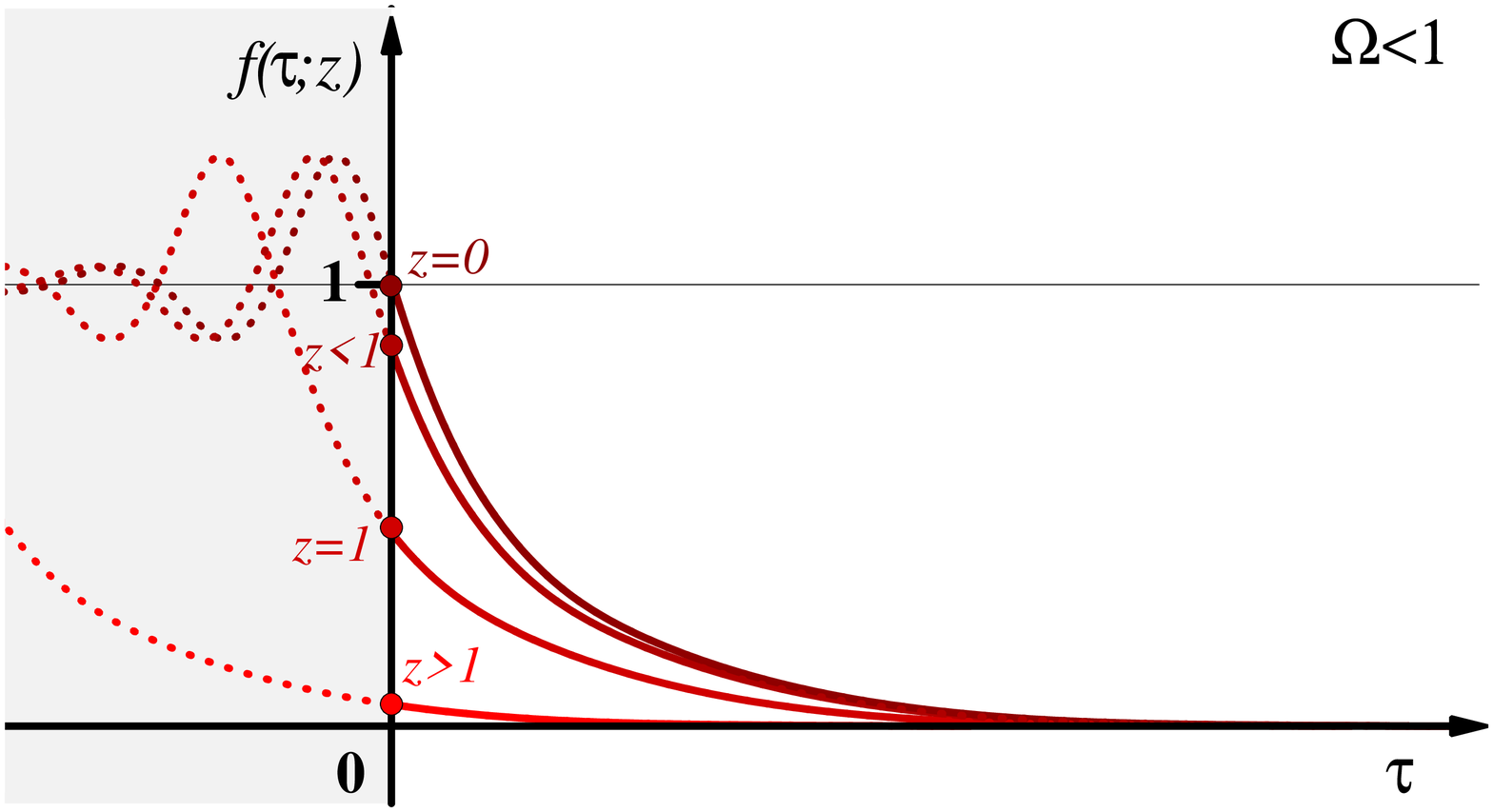}
\caption{\label{fig:fO} \small \it Schematic plots of the solutions $f(\tau;z)$ as a function of $\tau$ for different
choices of the boundary condition $z$ and for $\Omega>1$ (top) and $\Omega<1$ (bottom). For $\Omega>1$ when $z\to0$
the solution approaches $f(\tau;0)=1$ for every finite $\tau$'s---the probability to end inflation globally is 1; 
for $\Omega<1$ in the same limit the solution approaches a non-trivial function with $f(\tau;0)\ll1$ 
for large $\tau$---the probability to end inflation globally is small.}
\end{center}
\end{figure}

  We illustrate the behavior of the solutions in the two different regimes in fig.~\ref{fig:fO}.
  This plot makes very explicit the transition to the eternal regime at $\Omega=1$.
  At $\Omega>1$ by taking the limit $s_0\to 1$ one gets $f(\tau;0)=1$ for every $\tau$ so that the extinction 
  probability is $P_{\rm ext}=1$. On the other hand, in the same limit at $\Omega<1$ one obtains a non-trivial function $f(\tau;0)$
  that leaves the origin in a time of order $1/\sqrt{1-\Omega}$, therefore there is a non-vanishing probability 
  to inflate forever, $P_{\rm ext}=1-f(\tau;0)$.

In the following subsections we will present estimates for $\rho(V,\tau)$ in different regimes
and discuss finite-barrier effects. However, before entering the analysis of the probability distribution 
itself, we would like to discuss first its moments, which can be derived exactly
in a straightforward way from the master differential equation (\ref{eq:diffeq}).

\subsection{Moments of the probability distribution and critical points} \label{sec:moments}
In spite of the fact that the phase transition happens at the critical value $\Omega=1$,
in \cite{Creminelli:2008es} it was shown that, in the infinite barrier limit,
the moments of the probability distribution start diverging at different values
of $\Omega$: the higher the moment, the higher the value of $\Omega$ at which they diverge. 
On the other hand, in the finite barrier case all the moments diverge at the critical value $\Omega=1$. 
In the first part of this section, we will rederive these results in a rather quick and different way,
obtaining also the general formula for the critical value $\Omega$ at which each moment diverge. 
Then, in the second part, we will describe a procedure derived from the master differential equation~(\ref{eq:diffeq}) 
that allows us to easily compute the explicit value of each moment as a function of $\Omega$.

As shown in the previous section from eq.~(\ref{eq:momentsdef}), 
it is possible to extract the moments of the probability distribution $\rho(V,\tau)$ directly from $f(\tau;z)$
without the need of performing explicitly the anti-Laplace transform.
The $n$-th moment is indeed proportional to the $n$-th derivative of $f(\tau;z)$ with
respect to $z$, at $z=0$. Divergences in the moments thus correspond to a
non-analyticity of $f(\tau;z)$ at $z=0$. Therefore it is enough to study the solution
of the differential equation near the point $z=0$.
Notice that, before  entering the eternal regime, at $\Omega>1$, for every finite fixed value of $\tau$ and for smaller and smaller values of $z$, the solution 
is better and better described by the linear approximation of eq.~(\ref{eq:flinintro}) (see fig.~\ref{fig:fO}), 
which we conveniently rewrite here as
\be
f_{\rm lin}(\tau;z)=1-e^{\omega_-(\tau+\tau_0)}-\sigma e^{\omega_+(\tau +\tau_0)} \,,
\ee
where $$\omega_{\pm}\equiv\Oh\pm\sqrt{\Omega-1}$$ and $\sigma$ and $\tau_0$ are 
the two constant of integration.  In the infinite barrier case 
the boundary condition (\ref{eq:bc2}) corresponds to requiring that the solution stops 
on top of the hill ($f=0$) in an infinite time. Notice that since both
this boundary condition and the differential equation are invariant under shifts of $\tau$
we are left with a one-parameter family of solutions that are related by a shift in $\tau$, 
i.e. the parameter $\tau_0$. The latter is fixed by imposing the boundary condition
(\ref{eq:bc1}), namely
\be \label{eq:lin_bound}
f_{\rm lin}(0;z)=1-e^{\omega_-\tau_0 }-\sigma e^{\omega_+\tau_0}=e^{-z}\,.
\ee
This means that all the dependence on $z$ is in the parameter $\tau_0$,
while $\sigma$ is fixed by the boundary condition at infinity and is independent of $z$.
Using eq.~(\ref{eq:lin_bound}) the solution can be rewritten as 
\bea\label{eq:f_lin_refined}
f_{\rm lin}(\tau;z)&=&1-\l(1-e^{-z}\r)e^{\omega_-\tau}-\sigma e^{\omega_+ \tau_0}\l(e^{\omega_+\tau}-e^{\omega_-\tau}\r) \nn \\ 
&\simeq& 1-z e^{\omega_-\tau}-\sigma z^{\omega_+^2} e^{\omega_+\tau} \ ,
\eea
where in the second line we have used the approximate solution $\tau_0\approx \log(z)/\omega_-$ from eq.~(\ref{eq:lin_bound}) 
and dropped a subleading exponent in the last term. 

Notice also that the last term, in general, is not analytic in $z=0$. 
Indeed if we calculate the $n$-th derivative of $f_{\rm lin}(\tau;z)$ with respect to $z$ we get
\be
f^{(n)}_{\rm lin}(\tau;z)\sim z^{\omega_+^2-n} e^{\omega_+\tau}
\ee 
so that the moment $\la V^n\ra$ starts diverging when $\omega_+^2$ becomes smaller than $n$,
\emph{i.e.} at
\be \label{eq:wcrit}
\Omega=\frac{(n+1)^2}{4n}\,.
\ee
For instance, in the case of the variance ($n=2$), we get $\Omega=9/8$ as critical value,
in perfect agreement with the lengthy calculation of \cite{Creminelli:2008es} that used directly the inflaton stochastic  equations.
In the presence of a finite barrier this argument breaks down because now also $\sigma$
depends on $z$ and the analytic structure around the origin changes. 
In fact, for small enough values of $z$, $f(\tau;z)$ is still well described by the linear approximation around $f=1$ even at the barrier $\tau_b$.  It is not difficult then to solve the differential equation in the linearized limit with 
a finite barrier. What we get in this case is
\be
f_{\rm lin}(\tau;z)= 1-(1-e^{-z})\frac{e^{\omega_+ \tau+\omega_- \tau_b}-\omega_+^2 e^{\omega_-\tau +\omega_+ \tau_b}}
{e^{\omega_- \tau_b}-\omega_+^2 e^{\omega_+ \tau_b}}\,,
\ee
which is analytic in $z=0$ for all values of $\Omega$, implying that all the moments converge for $\Omega>1$.
Although we expect the linearized approximation to work better and better
as $z\to0$, we cannot be sure yet whether non-analytic terms may arise from subleading non-linear corrections. Still this argument suggests that the finite barrier case behaves differently than the infinite barrier case.

\begin{figure}[t!]
\begin{center}
\includegraphics[width=9cm]{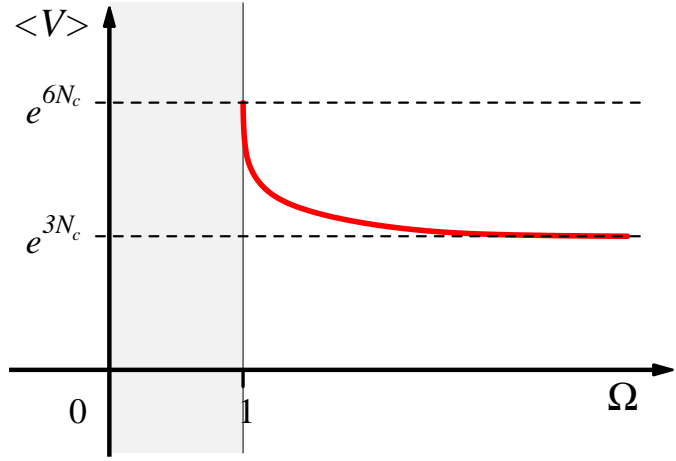}
\caption{\label{fig:Nav} \small \it Average volume as a function of $\Omega$, with $N_c\equiv\frac{2\pi^2}{\Omega}\frac{\phi}{H}=\frac{\tau}{6\sqrt{\Omega}}$.}
\end{center}
\end{figure}

As a proof of this statement we will now present a method to derive exactly all the moments.
Indeed, even though we are not able to solve analytically the non-linear differential equation (\ref{eq:diffeq}),
the equations for the moments are linear and can be solved explicitly.
They can be obtained by simply deriving $n$ times the eq.~(\ref{eq:diffeq}) 
with respect to $z$ at $z=0$.
For example by deriving (\ref{eq:diffeq}) once with respect to $z$ we get
\be \label{eq:diffav}
\ddot f'-2\Oh \dot f'+ f'+f'\log f=0\,,
\ee
where ``dots'' represent derivatives with respect to $\tau$ and ``~$'$~'' with respect to $z$.
Since for $z=0$ $f=1$ we get a linear differential equation for $f'_0\equiv f'(\tau;0)=-\la V \ra$ 
with solution
\be \label{eq:solav}
f'_0=A e^{\omega_+ \tau}+B e^{\omega_- \tau}\,.
\ee
The constants of integration $A$ and $B$ can be fixed using the derivative of 
the boundary conditions (\ref{eq:bc1}) and (\ref{eq:bc2}), namely
\be \label{eq:bcav}
f'_0(0)=-1\,,\qquad \dot f'_0(\tau_b)=0\,. 
\ee
This way we get
\be
\la V\ra=-f'_0(\tau) = \frac{e^{\omega_+ \tau+\omega_- \tau_b}-\omega_+^2 e^{\omega_-\tau +\omega_+ \tau_b}}
{e^{\omega_- \tau_b}-\omega_+^2 e^{\omega_+ \tau_b}}\,,
\ee
which in the large $\tau_b$ limit gives (see fig.~(\ref{fig:Nav}))
\be \label{eq:Vav1}
\lim_{\tau_b\to \infty} \la V\ra = e^{\omega_- \tau}=e^{3 N_c \frac{2}{1+\sqrt{1-1/\Omega}}}\,.
\ee
This results nicely agrees with the explicit calculation from the inflaton equation (see the appendix)
and with the result from the probability distribution $\rho(V,\tau)$ that we will derive in the 
next sections. Notice that for large $\Omega$ one recovers the classical limit for the average 
volume $V_c=e^{3N_c}$. With very little effort eq.~(\ref{eq:diffeq}) gave us the formula for the average
volume in both finite and infinite barrier cases.

Roughly with the same amount of work we can obtain also the expression for any higher moment.
By deriving eq.~(\ref{eq:diffav}) with respect to $z$ a second time we obtain
\be
\ddot f''-2\Oh \dot f''+ f''+ f''\log f+\frac{f'^2}{f}=0\,,
\ee
which again gives a linear differential equation at $z=0$,
\be \label{eq:diffvar}
\ddot f''_0-2\Oh \dot f''_0+ f''_0= -f'^2_0\,,
\ee
this time with a non-homogeneous source term. The latter however is an exponential, therefore
the differential equation can be easily solved analytically.
Imposing the boundary conditions
\be
f''_0(0)=1\,, \qquad \dot f''_0(\tau_b)=0\,,
\ee
the result for the second moment is 
\bea \label{eq:V2barr}
\la V^2 \ra &=& f''_0(\tau)=
\frac{\omega_+^6 e^{\frac{2 \tau}{\omega_+}+2 \tau_b \omega_+} }{\l(\omega_+^2-2\r)\l(e^{\tau_b/\omega_+}-e^{\tau_b \omega_+} \omega_+^2\r)^2}
-\frac{2\omega_+^4 e^{2 \tau_b \omega_+} \l(e^{\frac{\tau_b}{\omega_+}+\tau \omega_+}-e^{\frac{\tau}{\omega_+}+\tau_b \omega_+} \omega_+^2\r) }{\l(\omega_+^2-2\r)\l(e^{\tau_b/\omega_+}-e^{\tau_b \omega_+} \omega_+^2\r)^3}
\nn \\ &&-\frac{4\omega_+^2 e^{\omega_+ \tau_b+\frac{\tau_b}{\omega_+}} \l(e^{\frac{\tau_b}{\omega_+}+\tau \omega_+}-e^{\frac{\tau}{\omega_+}+\tau_b \omega_+} \omega_+^2\r) }{\l(e^{\tau_b/\omega_+}-e^{\tau_b \omega_+} \omega_+^2\r)^3}
+\frac{2 \omega_+^2e^{\frac{2 \tau_b}{\omega_+}} \l(e^{\frac{\tau_b}{\omega_+}+\tau \omega_+}-e^{\frac{\tau}{\omega_+}+\tau_b \omega_+} \omega_+^2\r) }{\l(2 \omega_+^2-1\r) \l(e^{\tau_b/\omega_+}-e^{\tau_b \omega_+} \omega_+^2\r)^3}
\nn \\ && +\frac{8 \omega_+^2 e^{2 \omega_+ \tau_b+\frac{2 \tau_b}{\omega_+}} \l(e^{\tau \omega_+} -e^{\tau/\omega_+}\r) \l(\omega_+^2-1\r)^2 \l(\omega_+^2+1\r)}{\l(e^{\tau_b/\omega_+}-e^{\tau_b \omega_+} \omega_+^2\r)^3 \l(2 \omega_+^4-5 \omega_+^2+2\r)}
+\frac{2 \omega_+^2 e^{\omega_+ \tau+\frac{\tau}{\omega_+}+\tau_b \omega_++\frac{\tau_b}{\omega_+}} }{\l(e^{\tau_b/\omega_+}-e^{\tau_b \omega_+} \omega_+^2\r)^2}
\nn \\&& -\frac{e^{\frac{2 \tau_b}{\omega_+}+2 \tau \omega_+}}{\l(2 \omega_+^2-1\r) \l(e^{\tau_b/\omega_+}-e^{\tau_b \omega_+} \omega_+^2\r)^2}\,,
\eea
where the length of the expression indicates how non-trivial it would have been to obtain this result
directly from the inflaton equation.
In the large barrier limit the asymptotic form of eq.~(\ref{eq:V2barr}) for $\Omega>1$ reads
\be
\la V^2 \ra \ \stackrel{\tau_b\gg1}{\longrightarrow}\ \frac{\omega_+^2}{\omega_+^2-2}\l(1-2\frac{e^{-\omega_- \tau}}{\omega_+^2}\r) 
e^{2\omega_- \tau}+\frac{8(\omega_+^2-1)^2(\omega_+^2+1)}{\omega_+^4 (2\omega_+-1)(2-\omega_+^2)}
e^{-(\omega_+^2-2)\omega_-\tau_b+\omega_+ \tau}\,,
\ee
the divergence at $\omega_+^2=2$ (i.e. $\Omega=9/8$) is manifest, in particular the last term 
vanishes for $\omega_+^2>2$ in the limit $\tau_b\to\infty$, while it explodes for  $\omega_+^2\leq2$.
It is less manifest from eq.~(\ref{eq:V2barr}) the fact that, with a finite barrier, there is no divergence;
for this propose we give here the expression for eq.~(\ref{eq:V2barr}) at $\Omega=9/8$
\be
\l.\la V^2 \ra \r|_{\Omega=9/8}=\sqrt{2}\tau_b (1-e^{-\tau/\sqrt2})e^{\sqrt2 \tau}+\dots\ ,
\ee
which shows that, up to sub-dominant terms in the large barrier limit (the dots),
the result is finite but linear in $\tau_b$, explaining the divergence in the infinite barrier case.

We could keep going calculating higher moments, indeed for the $n$-th moment
we have just to solve the following linear differential equation
\be \label{eq:diffn}
\ddot f^{(n)}_0-2\Oh \dot f^{(n)}_0+f^{(n)}_0=J^{(n)}\,,
\ee
\be f^{(n)}_0(0)=(-1)^n\,,  \ee
\be \dot f^{(n)}_0(\tau_b)=0\,, \ee
where the source $J^{(n)}$ is a polynomium of degree $n$ of the lower moments ($f^{(k)}_0$ with $k<n$)
\be
J^{(n)}=\l.\de_z^{n} [f(\log f-1)]\r|_{z=0}\,,
\ee
which will then be a sum of exponentials of the form $e^{k\omega_\pm \tau}$ at most of degree $k=n$.

Iterating the analysis it is possible to check that, in the infinite barrier limit,
the critical value of $\Omega$ where the $n$-th moment diverges perfectly agrees 
with eq.~(\ref{eq:wcrit}), while for finite barriers the moments converge for every $\Omega>1$.

\subsection{$\Omega\gg1$: Classical limit} \label{sec:classical}
Let us now find approximations for the probability distribution $\rho(V)$ in different regimes by directly 
using the Laplace transform formula (\ref{eq:rhoV}). 
The first case we will study is the limit $\Omega\gg1$, which corresponds to the conventional
slow-roll inflation, far from the eternal regime. As expected, 
we will see explicitly that in this case the volume probability is sharply peaked around the volume corresponding to  the classical  inflaton trajectory.
The main reason to start with this case is that we will be able to find an explicit 
solution to the mechanical problem (\ref{eq:diffeq}).
This will help us to develop an intuition on how to proceed also for generic 
values of $\Omega$, where such a solution is absent.

To analyze the large $\Omega$ limit it is convenient to 
rescale the time variable $\tau$ as
\be
\tau=2\Oh{\tilde\tau}\,.
\ee
The new time variable 
${\tilde\tau}$ measures the number of classical $e$-foldings $N_c$,
\be
{\tilde\tau}=3N_c\,.
\ee
Then the mechanical equation (\ref{eq:diffeq}) takes the following form
\be \label{eq:gdiffeq}
\frac{1}{4\Omega}\frac{\de^2 f}{\de {\tilde \tau}^2}-\frac{\de{f}}{\de\tilde \tau}+f\log{f}=0\,.
\ee
It is useful to present $f$ in the exponential form
\be
f=e^{-g}\,,
\ee
then the function $g$ satisfies
\be
\label{eq:gtau}
\frac{1}{4\Omega}\l[\frac{\de^2 g}{\de{\tilde\tau}^2}-\l(\frac{\de g}{\de\tilde\tau}\r)^2\r]-\frac{\de g}{\de \tilde \tau}+g=0\,.
\ee
The most straightforward way to do the large $\Omega$ expansion would be,
as a zeroth order approximation, to drop the first two terms in this equation. This would give
\be\label{eq:classical}
g=e^{{\tilde\tau}+{\tilde\tau}_0}\,,
\ee
where ${\tilde\tau}_0$ is an integration constant. 
This translates into 
\be
f=e^{-z\,e^{\tilde\tau}}\,,
\ee
when we fix ${\tilde\tau}_0$ to match the boundary conditions for $f$.
The corresponding probability distribution that we obtain from eq.~(\ref{eq:rhoV}) is
\be
\rho(V,\tau)=\delta(V-e^{3N_{c}})\,,
\ee
i.e. the inflaton follows the classical trajectory,
exactly what expected in the classical limit where quantum fluctuations can be neglected.
However,  there is a problem to use this solution as a basis for a systematic expansion around $1/\Omega=0$ because 
the corrections due to the dropped terms in (\ref{eq:gtau}) are not always small.
Indeed, for ${\tilde\tau}\gg\log\Omega$ one has ${\dot g}^2/\Omega\gg g, \dot g$ in this case. 
We can get around this problem by keeping this dangerous term in  (\ref{eq:gtau}), and dropping only the very first one
(similarly to the WKB approximation).
Now we have the following equation
 \be
\label{eq:gtauapprox}
\frac{1}{4\Omega}\l(\frac{\de g}{\de\tilde\tau}\r)^2+\frac{\de g}{\de \tilde \tau}-g=0
\ee
that after integration gives $g$ as a solution of the following algebraic equation,
\be
\label{productlog}
G e^G=e^{{\tilde\tau}+{\tilde\tau}_0}\,,
\ee
where
\[
G=\sqrt{1+\frac{g}{\Omega}}-1\,.
\]
For ${g/\Omega}\ll 1$ this gives back the previous result (\ref{eq:classical}), while in the opposite limit
${g/\Omega}\gg 1$ one gets 
\be
\label{largeg}
g=\Omega({\tilde\tau}+{\tilde\tau}_0)^2\,.
\ee
Importantly, for this solution the dropped term $\ddot g/\Omega$ is small compared to the other 
ones in (\ref{eq:gtau}) in both limits as long as $\Omega |G+1|\gg 1$.
Therefore this solution provides a basis for a consistent $1/\Omega$ expansion which is valid everywhere apart in the small region $|G+1|\lesssim 1/\Omega$.

\begin{figure}[t!]
\begin{center}
\includegraphics[width=17.5cm]{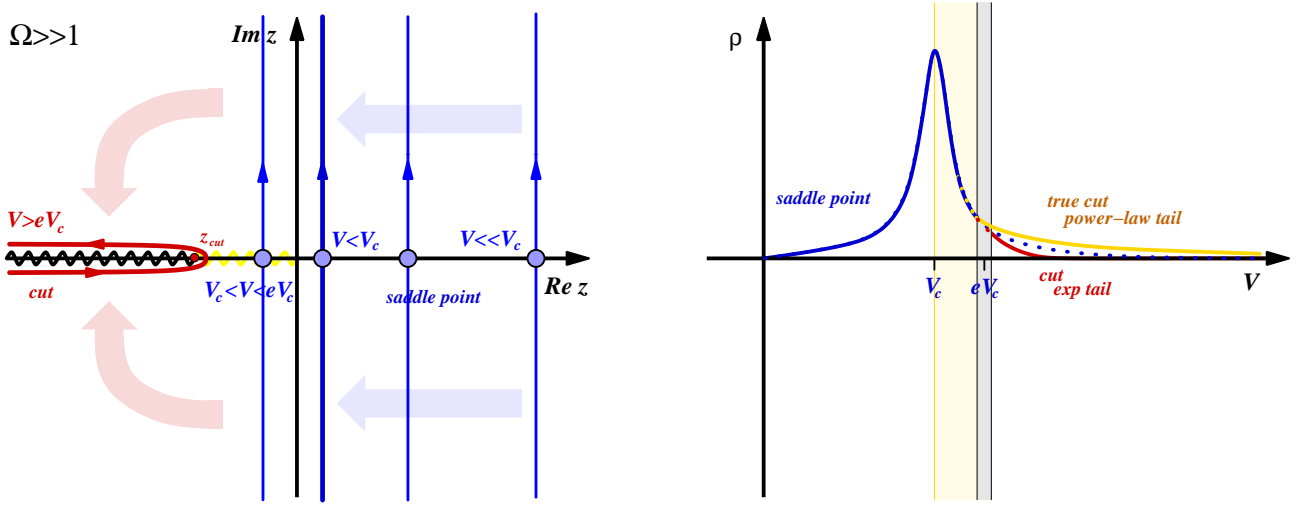}
\caption{\label{fig:zrho-Wlarge} \small \it Contour of the anti-Laplace transform integral in the $z$-plane \emph{(on the left)}
and probability distribution of the volume \emph{(on the right)} for $\Omega\gg1$. For small volumes the integral can be solved via saddle-point approximation
and gives a gaussian-like distribution \emph{(in blue)}. Near $V=e V_c$ the saddle point hits the cut, the contour of integration can be deformed
around the cut and for $V>eV_c$ the distribution becomes exponentially small \emph{(in red)}. The large $\Omega$ limit does not capture
the cut between $z_{cut}$ and $z=0$, which deforms the tail of the distribution 
(from a value of $V\in (V_c,e V_c)$ where $\rho(V)\approx e^{-\Omega}$)
making it power-like in the volume \emph{(in yellow)}.}
\end{center}
\end{figure}

Let us see what probability distribution one gets from the above solution.
We can find ${\tilde\tau}_0(z)$ by imposing the boundary condition\be
f(0;z)=e^{-z}=e^{-g({\tilde\tau}=0)}\,,
\ee
which gives
\bea
\label{tau0}
e^{{\tilde\tau}_0(x)}=Ae^A\,,
\eea
where we have defined
\be \label{eq:Ax}
A\equiv \sqrt{1+x}-1\,, \quad \quad x\equiv{z\over\Omega}\,.
\ee
Then expression (\ref{eq:rhoV}) for the probability distribution takes the following form
\be
\label{eq:rhoOmega}
\rho(V,\tau)=\frac{\Omega}{2\pi i}\int_{0^+-i\infty}^{0^++i\infty}dx\, e^{-\Omega\l( G^2+2G-xV\r)}\;,
\ee
where $G$ is a function of $x$ via eqs.~(\ref{productlog}), (\ref{tau0}) and (\ref{eq:Ax}).
We can try to evaluate this integral using the saddle point approximation 
(we will check later whether this is a good approximation). The result reads
\be
\rho(V,\tau)\approx \frac{\Omega}{\sqrt{2\pi |S''(x_0)|}}e^{-S(x_0)}\,,
\ee
where $S(x)$ is given by
\be
S(x)=\Omega\l( G^2+2G-xV\r)\,,
\ee
and $x_0$ is the saddle point satisfying the equation
\be
\label{saddlefirst}
S'(x_0)=\l.\Omega\l(2(1+G)G' \frac{\de {\tilde\tau}_0}{\de x}-V\r)\r|_{x=x_0}=0\,.
\ee
By taking the derivative of  (\ref{productlog}), and plugging in the resulting expression for $G'$
into the saddle point condition (\ref{saddlefirst}) one gets
\be
\label{VA}
G=V A
\ee
at the saddle point. If we substitute this expression back in eq.~(\ref{productlog})  we get
\bea
Ge^G&=&A V e^{A V}=A e^{{\tilde\tau}+A}\,, \\
&\Rightarrow&A= \frac{1}{V-1}\log\l(\frac{e^{3N_c}}{V}\r)\;.
\label{Aeq}
\eea
Note that  for $V < V_c\equiv e^{3N_c}$ the value of $A$ is positive and the saddle point is at  real and positive $z$ (see fig.~\ref{fig:zrho-Wlarge}).
On the other hand for $V > V_c$ the value of $A$ is negative and the saddle point is located at
 negative $z$ \footnote{This implies that for $V> V_c$ the saddle point corresponds to a value for the boundary condition $z$ that is outside of the physical region $z>0$. We are using the analytic continuation of the solution, as it is implicit in the definition of the inverse-Laplace transform of eq.~(\ref{eq:rhoOmega}).}. 
 To understand when the saddle point approximation is applicable for calculating the integral (\ref{eq:rhoOmega}) it is important to note that our approximate solution has a branch cut starting
 at the point $z_{cut}\approx -2\Omega e^{-\tilde\tau-1}$, where $G=-1$ (equivalently, $g=-\Omega$). 
 An easy way to see this is to use eq.~(\ref{eq:gtauapprox}) as an equation for $\partial g/\partial \tilde\tau$, namely
 \be
 \frac{\d g}{\d \tilde\tau}=2\Omega\left(-1+\sqrt{1+\frac{g}{\Omega}}\right)\ .
 \ee
We see that at $g=-\Omega$
there is a discontinuity in the imaginary part of $\d g/\d \tilde\tau$ along the real axis in the $g$-plane at $g<-\Omega$, leading to a cut
for $g$ as a function of $z$ at real $z<z_{cut}$. 

From (\ref{VA}) and (\ref{Aeq}) we find that the saddle point hits the cut at $ V\simeq e V_c$.
As long as the saddle point does not hit the cut, {\it i.e.} for $V\lesssim e V_c$, we can use the saddle point approximation to perform the integral. Plugging the solution in eqs.~(\ref{VA}) and (\ref{Aeq}) back into $S$ we get
\bea
S(x_0)&=&\Omega\frac{(V-1)}{V} G^2=\Omega\frac{V}{V-1}\l[\log\l(\frac{V}{V_c}\r)\r]^2\,, \\
S''(x_0)&=&{\textstyle\frac{\Omega (V-1)}{2}\l(1-\frac{\log\l(\frac{V}{V_c}\r)}{V-1}\r)^{-1}\l(\frac1V-\frac{\log\l(\frac{V}{V_c}\r)}{V-1}\r)^{-1}}\,,
\eea
and thus
\be \label{eq:rhoVNN}
\rho(V,\tau)\approx   
{\cal N} 
e^{-\Omega\frac{V}{V-1}\l[\log\l(\frac{V}{V_c}\r)\r]^2}\,, \quad\quad V\lesssim e V_c \ ,
\ee
where the prefactor ${\cal N}$ is equal to
\be
{\cal N}= {\textstyle \l|\frac{\Omega}{\pi\,V(V-1)}\l(1-\frac{1}{V-1}\log{\frac{V}{V_c}}\r)\l(1-\frac{V}{V-1}\log{\frac{V}{V_c}}\r)\r|^{1/2}}\,. \nn
\ee
Notice that $S''(x_0)$ is large for exponentially large volumes, making our saddle point approximation justified. 

For large volumes ($V\gg 1$) the formula (\ref{eq:rhoVNN}) simplifies to
\be
\rho(V,\tau)\sim {\sqrt{\frac{\Omega}{\pi}\l|\log\frac{e V_c}{V}\r|}}\frac{1}{V}\,e^{-\Omega\l(\log\frac{V}{V_c}\r)^2}\,,  \quad\quad V\lesssim e V_c \ ,
\ee
which we can also rewrite as the probability distribution to have $N$ $e$-foldings
\be
\tilde\rho(N,N_c)=3V\rho(V,\tau)\sim3\sqrt{\frac{\Omega|3N-3N_c-1|}{\pi}}\,
e^{-\Omega\l(3N-3N_c\r)^2}\, ,  \quad\quad N\lesssim  N_c \, .
\ee
This distribution is a gaussian centered around
the classical number of $e$-foldings $N_c$ (see fig.~\ref{fig:zrho-Wlarge}). The spread is of order $1/\Oh$ and goes to zero
as $\Omega$ goes to infinity reproducing the $\delta$-function of the classical limit.

When the volume becomes approximately $e V_c$ the saddle point reaches the cut and we cannot perform the saddle point approximation anymore. In this regime,  we can still close the contour of integration on the left around the cut (see fig.~\ref{fig:zrho-Wlarge}), and perform the integral along the discontinuity. We obtain
\bea\label{eq:int_cut1}
\rho(V,\tilde\tau)=\frac{1}{2\pi i}\int_{|z_{\rm cut}|}^{+\infty}d |z|\; 2 i\, {\rm Im}[f(\tilde\tau;-|z|)] e^{- V |z|}\ .
\eea 
It is straightforward to verify the $|f(\tilde\tau;z)|$ does not grow faster than $e^{|z|}$ at large $z$. 
Therefore, at large volumes the integral above is dominated by values of $z$ very close to the cut, with a spread of the order $1/V$. We can thus write approximately
\be
\label{fake}
\rho(V,\tilde\tau)\sim e^{z_{\rm cut} V}\sim e^{-2\frac{\Omega}{e}V/ V_c} \, , \quad \quad V\gtrsim e V_c \ ,
\ee
where we have ignored power corrections in the volume.
We see that  the probability distribution has an exponential tail in $V$ that starts many standard deviations away from the average $V_c$. This result confirms that after the saddle point hits the cut 
one cannot use it any longer to evaluate the Laplace transform. Indeed, if one keeps 
using the saddle point approximation blindly one would obtain the gaussian behavior for the probability distribution
up to arbitrary large volumes, in contradiction to (\ref{fake}).


However, there is a problem with the behavior (\ref{fake}) as well.
Namely, this result disagrees with our analysis in section~\ref{sec:moments}, where we found that high
enough moments of the volume distribution diverge at any value of $\Omega$. Related to this, we proved
there that the function $f$ has a cut starting at $z=0$, while here we are finding the origin of the cut at
$z=z_{cut}<0$.

The origin of this discrepancy is the non-commutativity of the large $\Omega$ limit and the large volume limit. One indication
of the problem is that the large $\Omega$ expansion breaks in the vicinity of $G=-1$---precisely where the cut of the 
approximate solution starts.
Even more relevant is the following observation. The leading non-analytic term in  (\ref{eq:f_lin_refined}),
 that gives rise to the cut
starting at $z=0$, is proportional to $ z^{4\Omega}$. At small $z$ this term is non-perturbatively small in
the large $\Omega$ expansion.
However, it  dominates the behavior of $\rho(V)$ at large volumes and gives rise to the power-law
tail proportional to $V^{-4\Omega}$. Our approximate solution misses the corresponding 
 part of the cut between 0 and $z_{cut}$.
 Note, however, that  this part of the cut
becomes important only at  volumes much larger than the average, where the probability is exponentially
suppressed in $\Omega$, $\rho(V)\propto e^{-\Omega}$. Consequently, our approximate solution correctly reproduces the shape of $\rho(V)$ up to $V\lesssim eV_c$. At larger volumes instead of the exponential 
behavior  (\ref{fake}) one gets the power-law tail. In the next section we will discuss this tail 
in more details for general $\Omega>1$.

\subsection{$\Omega\ge1$: Approaching the phase transition} \label{sec:Wgen}
Let us now reconstruct the probability distribution of the volume for generic $\Omega\ge1$.
Unlike in the previous case, we do not have any small parameter that allows us to find
an approximate full solution to the mechanical problem (\ref{eq:diffeq}). However we will  be able to find
an approximate form of the probability distribution $\rho(V)$ practically at all $V$.

In order to do this, we notice that we can solve the differential equation (\ref{eq:diffeq}) (and equivalently eq.~(\ref{eq:gdiffeq})) in two different regimes. When $f\simeq 1$ (equivalent to $g\ll1$), we can linearize the potential and obtain the solution 
\be\label{eq:f_lin2}
f_{\rm lin}(\tau;z)= 1-e^{\omega_- (\tau+\tau_0)}-\sigma e^{\omega_+(\tau+\tau_0)}\,,
\ee
where $\omega_\pm\equiv\Oh\pm\sqrt{\Omega-1}$ and $\sigma$ and $\tau_0$ are integration constants. For the linear approximation to hold, it is enough to impose that $\tau+\tau_0\ll -1$. Then, independently of the value of $\sigma$, for large enough  $\tau+\tau_0$  we can approximate the solution as
\be
f_{\rm lin}\approx 1-e^{\omega_-(\tau+\tau_0)}\,.
\ee
The second regime in which we can solve the differential equation is when $f\simeq 0$ (equivalent to $g\gg1$). 
In this regime the two dominant terms in (\ref{eq:gtau}) are those proportional to $g$ and $({\d g\over \d \tau})^2$;
by dropping the other terms one obtains
\be\label{eq:fg}
f\approx f_{g}= e^{-\frac{(\tau+\tau_1)^2}{4}}\,,
\ee
where $\tau_1$ is an integration constant. 
By plugging this solution back into the equation, one may check that this approximation indeed holds as long as $g\gg1$, i.e. for $|\tau+\tau_1|\gg1$.

Notice that the constants of integrations in both cases can be absorbed into a shift ($\tau_0$ or $\tau_1$) of the ``time" variable $\tau$. $\tau_0$ and $\tau_1$ are in general not equal---they differ by an unknown constant of order one set by the matching of the two solutions. However, later we will be interested in the large $(\tau+\tau_0)$ limit where  such a constant can be neglected and the $\tau_0$ and $\tau_1$ can be taken as equal.

Let us see now that this information is enough to reconstruct $\rho(V)$ almost for all $V$.
First of all,  we need to know how $f$ depends on $z$. From  section~\ref{sec:moments} (see eq.~(\ref{eq:f_lin_refined})), 
we know that $f(\tau;z)$ has a branch cut at $z=0$. We could perform the integral by closing the contour around the cut, or by using the saddle point approximation. Let us start by seeing if and where we can use the saddle point approximation.

For large volumes the saddle point
is expected to lie at small $z$ ($f\simeq1$), where the exponential suppression in (\ref{eq:rhoV}) is milder. 
In this case, near $\tau=0$, the linearized approximation holds, and we can use $f_{\rm lin}$ to relate $\tau_0$ to $z$.
 Next, in order to evaluate the integral  (\ref{eq:rhoV}), we need to impose that the saddle point belongs to one of the regions where our asymptotic solution works. Within the linearized regime there is no saddle point, so 
 we will check whether the saddle point  exists in the gaussian region at  $\tau+\tau_0 \gg1$.

\begin{figure}[t!]
\begin{center}
\includegraphics[width=17.5cm]{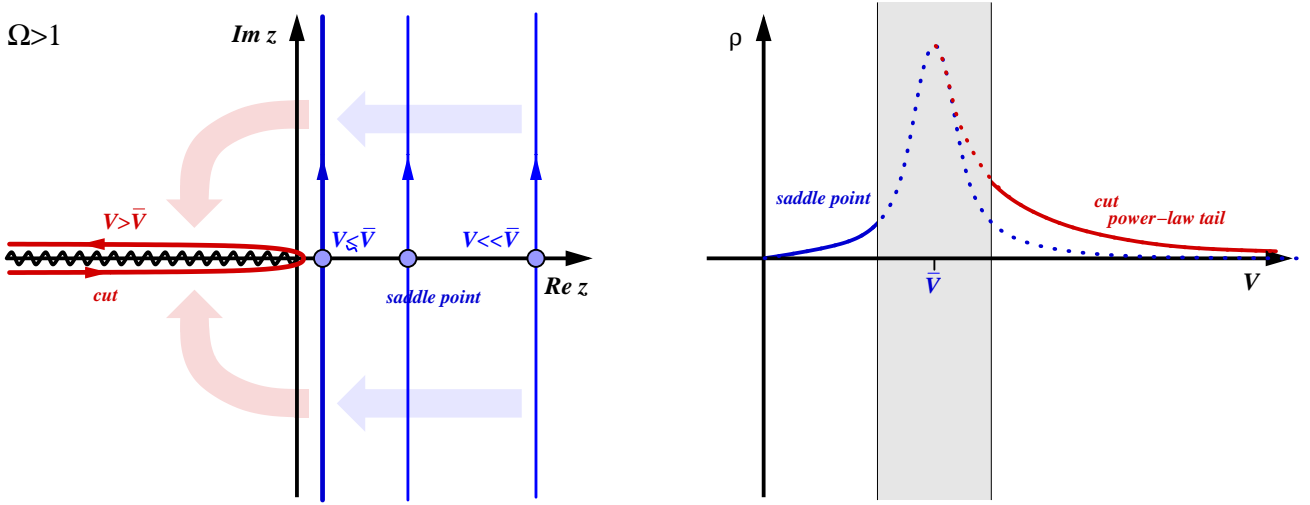}
\caption{\label{fig:zrho-Wgen} \small \it Contour of the anti-Laplace transform integral in the $z$-plane \emph{(on the left)}
and probability distribution of the volume \emph{(on the right)} for $\Omega\gtrsim1$. For small volumes the integral can be solved via saddle-point approximation
and gives a gaussian-like distribution \emph{(in blue)}. Around $V=\overline V$ \emph{(gray region)} the saddle point hit the cut, which starts at
$z=0$, and our approximations for the solution of the differential equation break down. 
At larger volumes there is no saddle point anymore but the contour of integration can be deformed
around the cut, giving a distribution tail that follows a power-law in the volume \emph{(in red)}.}
\end{center}
\end{figure}

Let us see how far the outlined procedure takes us and let us begin to implement it. Assuming that $z$ at the saddle point is small (we will check this assumption later), we can determine $\tau_0$ from the boundary condition
\bea\label{eq:lin_boundary}
e^{-z}=f_{\rm lin}(\tau=0;z)\approx 1-e^{\omega_-\tau_0}\,\quad\quad \Rightarrow\quad\quad
\tau_0\approx\frac1{\omega_-}\log z\,.
\eea
We can now substitute this value for $\tau_0$ in $f_g$ and perform the integral (\ref{eq:rhoV}) obtaining
 \be
\rho(V,\tau)\approx \frac{1}{\sqrt{2\pi |S''(z_0)|}}e^{-S(z_0)} \equiv {\cal N}e^{-S(z_0)}\,,
\ee
and $S(z)$ is given by
\be
S(z)=\frac14\l(\tau+\frac1{\omega_-}\log z\r)^2-V z\ .
\ee
The saddle point condition $S'(z_0)=0$ reads
\bea
\omega_- z_0 V=\frac12\l(\tau+\frac1{\omega_-}\log z_0\r)\,\quad\quad \Rightarrow \quad\quad
z_0\approx \frac{1}{2\omega_- V} \l [ \tau-\frac1{\omega_-}\log\l(\frac{2\omega_- V}{\tau}\r)\r]\,.
\eea
We see that for relatively small volumes $z_0$ is positive and small, which justifies our assumption to use the linearized limit to match $\tau_0$ with $z$. As $V$ grows, $z_0$ moves towards zero (see fig.~\ref{fig:zrho-Wgen}) and reaches the region where the gaussian approximation breaks when $V\simeq \overline V $, with $\overline V$ given by 
\be
\overline V \equiv e^{\omega_- \tau}= e^{3N_c \frac{2}{1+\sqrt{1-1/\Omega}}} \, .
\ee 
Even though we can not trust the gaussian approximation for $V\simeq \overline V$, we can try to follow the location of the saddle point, and we can see that it moves to negative values for $V\gtrsim \overline V$, reaching the location of the cut. This further justifies the approach we will take in the regime $V\gtrsim\overline V$, that is to do the integral along the cut.

For the moment instead let us concentrate on the regime of volumes $V\lesssim \overline V $, where we can apply the saddle point approximation.
Substituting the value of the saddle point back into $S(z_0)$ we get
\be
S(z_0)\approx \frac14\l(\tau-\frac{1}{\omega_-}\log V\r)^2=\Omega\l[3N_c-3N\l(\frac{1+\sqrt{1-\frac1\Omega}}2\r)\r]^2\ .
\ee
We  see that $\tau+\tau_0$ at the saddle point is large whenever $S(z_0)$ is large, 
i.e. at large $N$ and $N_c$ as long as $N\lesssim 2(\Omega-\sqrt{\Omega(\Omega-1)}) N_c$, (compatibly with the condition $V\lesssim \overline V $), so that in the same regions we can trust the use of $f_g$ for the saddle point.
The corresponding value of $S''(x_0)$ is
\be
S''(z_0) \approx \frac{2 V^2 (\sqrt{\Omega}-\sqrt{\Omega-1})^2}{\log\l(\frac{V}{\overline V}\r)}\ .
\ee
As in the large $\Omega$ limit, $S''(x_0)$ is large for large volumes.
So we conclude that the probability distributions for the volume after inflation for $V\lesssim \overline V$ and for generic $\Omega\ge1$ has the form
\be
\label{result}
\rho(V,\tau)\approx 
{\cal N}
e^{-\frac{1}{4}\Omega\left(1+\sqrt{1-\frac{1}{\Omega}}\right)^2\l[\log\left(\frac{V}{\overline V}\right)\r]^2}={\cal N} e^{-\Omega\l[\frac{3N}2\l(1+\sqrt{1-\frac1\Omega}\r) -3N_c\r]^2}\ ,\quad\quad V\lesssim \overline V \ ,
\ee
where in the last step we used that $3N=\log V$. 
We can trust this expression as long as $\log \overline V\gtrsim N\gg1$  (see fig.~\ref{fig:zrho-Wgen}). 

Let us now deal with the regime $V\gtrsim \overline V $. In this case, we have seen above that the saddle point $z_0$ enters the region very close to zero where we can not trust anymore the gaussian approximation. In section~\ref{sec:classical}, eq.~(\ref{eq:f_lin_refined}), we saw that $f(\tau;z)$ has a branch cut at the point $z=0$. Near the branch cut, for sufficiently small values of $z$, we can take the linear approximation $f_{\rm lin} (\tau;z) $, which reads
\bea\label{eq:f_lin_refined2}
f_{\rm lin}(\tau;z)&=&1-\left(1-e^{-z}\right)e^{\omega_-\tau}-\sigma e^{\omega_+ \tau_0}\left(e^{\omega_+\tau}-e^{\omega_-\tau}\right)\nn\\
&\simeq&1-z e^{\omega_-\tau}-\sigma z^{\frac{\omega_+}{\omega_-}} e^{\omega_+\tau} \, ,
\eea
where in the last line we have used the approximate solution $\tau_0\approx \log(z)/\omega_-$ of eq.~(\ref{eq:lin_boundary}). It is important to keep in mind that $\sigma$ does not depend on $z$, but it is only fixed by the boundary condition at $\tau\rightarrow +\infty$.

Since $\omega_+/\omega_-=\omega_+^2$ in general is not integer, from eq.~(\ref{eq:f_lin_refined2}) we see that $f_{\rm lin}(\tau;z)$ has a branch cut at $z=0$. In the $\Omega\gg 1$ case we did not see this cut starting at $z=0$ because, at large $\Omega$, $\omega_+/\omega_{-}\simeq 4\Omega$. At small $z$, this is a singularity that appears only non-perturbatively in $1/\Omega$ and could not be seen in a perturbative expansion in $1/\Omega$ as we did in the previous section.

We can now perform the integral of the discontinuity along the cut. As in section~\ref{sec:classical}, we have to compute the imaginary part of $f(\tau;z)$ along the cut and then integrate it
\bea\label{eq:int_cut2a}
\rho(V,\tau)=\frac{1}{2\pi i}\int_{0}^{+\infty}d |z|\; 2 i\, {\rm Im}[f(\tau;-|z|)] e^{- V |z|}\, .
\eea 
Since $f(\tau;z)e^{V z}$ rapidly decreases  when  the real part of $z$  is negative, 
at large enough volumes there is an interesting regime where the integral is dominated by small values of $z$, such that we can perform the integral using the linearized expression for $f(\tau;z)$. In this regime we have
\be
{\rm Im}[f_{\rm lin}(\tau;z)]_{\rm cut}\sim  e^{\omega_+\tau}z^\frac{\omega_+}{\omega_-}\ .
\ee
where we neglected order one coefficients.
By performing the integral (\ref{eq:int_cut2a}) we obtain
\be
\rho(V,\tau)\sim{1\over V}\left(\frac{\overline V}{V}\right)^{\frac{\omega_+}{\omega_-}}\quad {\rm for} \quad V\gtrsim \overline V\frac{\omega_+}{\omega_-}\,,
\ee
where we used that $e^{\omega_-\tau}=\overline V$.
The condition on the right, that determines how large the volume should be for this approximation to work,
 comes from imposing the validity of the linear approximation for $f(\tau;z)$. Indeed, the integral is dominated by values of $z$ around $z_s = \frac{\omega_+}{\omega_-}\frac{1}{V}$. For the linearized
 approximation to work
 $f_{\rm lin}(z_s;\tau)$ should be close to one. By plugging in the value for $z_s$ in the expression for $f_{\rm lin}$, we find
\be
f_{\rm lin}-1\simeq \frac{\omega_+}{\omega_-}\frac{\overline V}{V}+\sigma \left( \frac{\omega_+}{\omega_-}\frac{\overline V}{V}\right)^{\frac{\omega_+}{\omega_-}}\,,
\ee
which implies that the integral on the cut is well approximated by the integral of the linear solution for $V\gtrsim \frac{\omega_+}{\omega_-}\overline V$.

Similarly to the classical limit, the distribution for $V\lesssim \overline V $ is a gaussian in the number of $e$-folding centered at $\overline V$ with still a quite narrow width, of order one $e$-folding. 
Our method does not allow to reconstruct $\rho(V)$ in the vicinity of the average, $\overline V\lesssim V\lesssim \overline V \omega_+/\omega_-$. Note, however, that close to the eternal regime $ \omega_+/\omega_-$ is of order one, so that this range of volumes is not big. On the other hand at large $\Omega$,  when $ \omega_+/\omega_-$ is also large, we were able to find the probability distribution up to $V\simeq e\overline V$, where $\rho(V)$ was already exponentially suppressed as $e^{-\Omega}$.

For $V\gtrsim \overline V \omega_+/\omega_-$ the tail of the distribution in $V$ 
is power law,
$$\rho(V,\tau)\sim V^{-1-\frac{\omega_+}{\omega_-} }\sim V^{-\Omega\left(1+\sqrt{1-\frac{1}{\Omega}}\right)^2-1}\ .$$ 
This tail agrees with our results in section \ref{sec:moments} on the divergence of the multipoles (\ref{eq:wcrit}).

As a cross-check of these results note that they imply
that the average volume after inflation is given by
\be \label{eq:vavdiff}
\la V\ra \simeq e^{\frac{6N_c}{1+\sqrt{1-\frac1\Omega}}}=\overline V\,,
\ee
in agreement with the direct computations in section~\ref{sec:moments} (eq.~(\ref{eq:Vav1})) and in the appendix (eq.~(\ref{eq:vavinfl})).
We see that as $\Omega$ approaches the critical point the average number of $e$-foldings
shifts from $N_c$ in the limit $\Omega\to\infty$ to $2N_c$ in the $\Omega\to1$ limit.

\subsection{$\Omega\lesssim1$: Inside eternal inflation}
\label{sec:omegal1}

We now begin to explore the regime of eternal inflation. As $\Omega$ crosses 1, the solution of eq.~(\ref{eq:diffeq}) changes its form. The behavior of $f(\tau;z)$ around $f\simeq1$
is not overdumped anymore.
As discussed before, the normalization
of the probability distribution drops below 1 in this regime.

We would like to follow this transition carefully. To this purpose, we take $\Omega=1-\epsilon$ with $0<\epsilon\ll1$ and study the volume probability distribution  within the eternal inflation regime by expanding in $\epsilon$.
We will follow the same strategy as in the previous two subsections. We expect that the solution $f(\tau;z)$ has a branch cut in the complex $z$ plane, which allows us to perform the inverse-Laplace transform either with a saddle point approximation, if the saddle point is away from the cut, or along the cut itself.

If we decide to apply the saddle point approximation,  we can concentrate on large enough $\tau$, so that the saddle point lies in the region $f\to0$ where the solution is well approximated by
\be\label{eq:fg2}
f_g(\tau;z)=e^{-\frac{(\tau+\tau_0)^2}{4}}\,, \quad \quad \tau+\tau_0\gg 1\ ,
\ee
(we will check later  what is the corresponding range of volumes) and $\tau_0$ can be determined in terms of $z$ in the linearized regime (as long as $z\ll1$, as we will
check below). The linearized solution is now given by
\be\label{eq:flin_2}
f_{\rm lin}(\tau;z)=1-\sigma e^{\sqrt{\Omega}(\tau+\tau_0)}\cos\l(\sqrt{\Omega-1}(\tau+\tau_0)\r)\approx
1-\sigma e^{\tau+\tau_0}\cos\l(\eh(\tau+\tau_0)\r)\,.
\ee
Note the oscillatory behavior of  this linearized solution.
By rescaling the constant $\sigma$, the constant $\tau_0$ can be chosen equal to that in eq.~(\ref{eq:fg2}). $\tau_0$ will be fixed below 
in terms of the initial condition $z$; the constant $\sigma$ on the other hand is fixed by matching with the gaussian solution. Notice that, as before, $\sigma$ does not depend on $z$. We do not know the explicit value of $\sigma$, but we can argue that $\sigma\sim {\cal O}(1)$.  
Indeed, the gaussian solution breaks down when $\tau+\tau_0\sim {\cal O}(1)$, afterwards  the solution  will reach the linear regime in a time $\Delta\tau\sim{\cal O}(1)$; this means that the linear approximation breaks down when  $\tau+\tau_0\sim {\cal O}(1)$,  implying $\sigma\sim {\cal O}(1)$.



Let us now start the computation by fixing the relation between $\tau_0$ and $z$ via
\bea
e^{-z}=f_{\rm lin}(0;z)=1-\sigma e^{\tau_0}\cos\l(\eh \tau_0\r)\,\quad \Rightarrow \quad z\approx\sigma e^{\tau_0}\cos\l(\eh \tau_0\r)\,,\label{eq:solt0}
\eea
which is valid as long as $\tau_0\ll-1$.
Since the solution $f$ is constrained to be between 0 and 1, in particular 
we need  that for all $\tau$ in $0\leq\tau\lesssim -\tau_0$ 
\be
f_{\rm lin}(\tau;z)\le1\,,
\ee
which gives the lower bound
\be\label{eq:bt0}
\tau_0 \ge -\frac{\pi}{2\eh}\,.
\ee

Let us now study the analytic structure of $f(\tau;z)$. 
Continuity in $\epsilon$ suggests that $f(\tau;z)$ has a branch cut 
at large enough negative $z$ as before. As $\epsilon$ approaches zero the origin of this cut  $z_{cut}$
goes to $z=0$ as well. To follow how $z_{cut}$ depends on $\epsilon$ at small $\epsilon$ note
that eq.~(\ref{eq:flin_2}) implies  that $f_{\rm lin}(\tau;z)$ is a function of $z$ through its dependence on $\tau_0$, upon which it depends analytically. So, the only non-analyticity in $z$ of $f(\tau;z)$ can come from a non-analyticity of $\tau_0(z)$. The boundary condition (\ref{eq:solt0}) tells us that $z(\tau_0)$ is analytic. In inverting this relationship, therefore, the only non-analyticity can arise if $d z/d \tau_0$ vanishes at some value of $z$,
\bea
0&=&\frac{d z}{d\tau_0}=\frac{\sigma e^{\tau_0}\left(\cos(\sqrt{\epsilon}\tau_0)-\sqrt{\epsilon}\sin(\sqrt{\epsilon} \tau_0)\right)}{1-\sigma e^{\tau_0}\cos(\sqrt{\epsilon}\tau_0)}=0  \\ \nonumber
 &\Rightarrow& \cos(\sqrt{\epsilon}\tau_0)-\sqrt{\epsilon}\sin(\sqrt{\epsilon} \tau_0)=0 \\ 
 &\Rightarrow& \tau_0\simeq-\frac{\pi}{2\sqrt{\epsilon}}-1\ .
\eea
By plugging this value into (\ref{eq:solt0}), we obtain 
\be
z_{\rm cut}\simeq -\frac{\sigma\sqrt{\epsilon}}{e V_\epsilon}\ ,
\ee
where we defined 
\be
V_\epsilon\equiv e^\frac{\pi}{2\sqrt{\epsilon}}\ .
\ee

With this in mind, we can begin to evaluate the probability distribution using  the saddle point approximation. As in the previous section
\bea\label{eq:saddle_generic}
\rho(V,\tau)&\approx& \frac{1}{\sqrt{2\pi |S''(z_0)|}}e^{-S(z_0)}={\cal N}e^{-S(z_0)}\,,\\
S(z)&=&\frac14\l(\tau+\tau_0\r)^2-z V\,,
\eea
where the saddle point $z_0$ is determined by
\bea
S'(z_0)=\frac12\l[\tau+\tau_0(z_0)\r]\tau'_0(z_0)-V=0 \quad\quad\Rightarrow\quad
\quad V=\frac{(\tau+\tau_0)e^{-\tau_0}}{2\sigma \l[\cos(\eh \tau_0)-\eh\sin(\eh \tau_0)\r]}\,.\label{eq:Vt0}
\eea
The above relationship implies that for volumes that are large but smaller than $\overline V=e^\tau$, $\tau_0$ is large and negative, so that the linear approximation in (\ref{eq:solt0}) is justified. $\tau+\tau_0$ is also large and positive, so that we can trust the gaussian approximation for $f(\tau;z_0)$. As long as $\tau_0\gtrsim -\pi/(2\sqrt{\epsilon})=-\log V_\epsilon$, the denominator does not vanish, and $\tau_0$ is approximately given by
\be
\tau_0\approx-\log V\,.
\ee
The condition that $\tau_0\gtrsim- \log V_\epsilon$ then reads $V\lesssim V_\epsilon$. When plugged back into eq.~(\ref{eq:saddle_generic}) the above solution gives the same form for $\rho(V,\tau)$ 
as in the case $\Omega\ge1$, namely
\be\label{eq:fsol1}
\rho\approx 
{\cal N}e^{-\frac14\l(\tau-\log V\r)^2} \qquad {\rm for} \qquad 1\ll V\lesssim {\rm Min}\{\overline V,\, V_\epsilon\} \,,
\ee
where 
\be
{\cal N}=\frac{1+\sqrt{1-\frac1V}}{2\pi} \sqrt{\log\l(\frac{V}{\overline V}\r)} \ .
\ee

\begin{figure}[t!]
\begin{center}
\includegraphics[width=17.5cm]{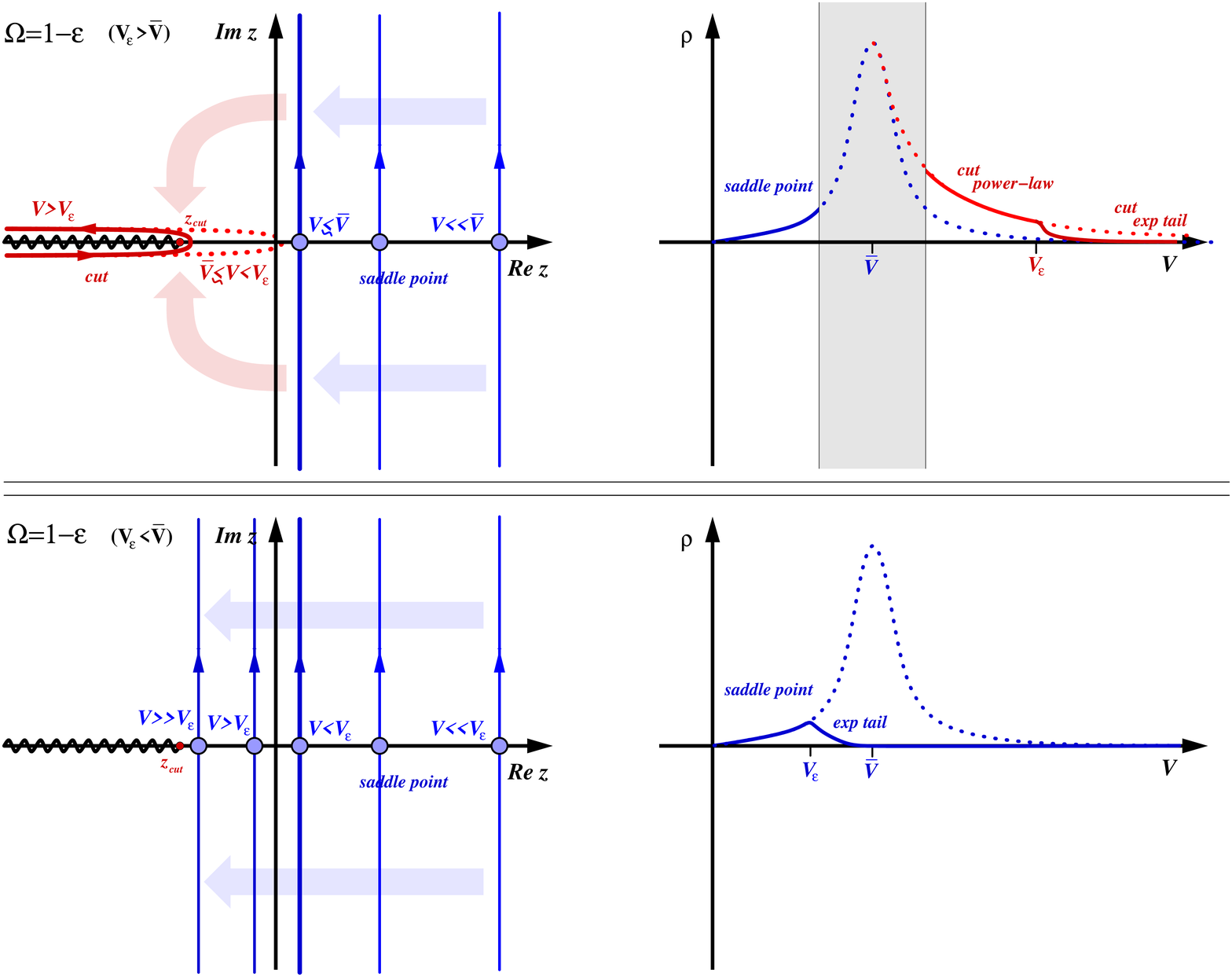}
\caption{\label{fig:zrho-Wl1} \small \it Contours of the anti-Laplace transform integrals in the $z$-plane \emph{(on the left)}
and probability distributions of the volume \emph{(on the right)} for $\Omega=1-\epsilon$.\newline
\mbox{}~~~ First case $\sqrt\epsilon<\pi/{(2\tau)}$ \emph{(first row)}: at small volumes the integral can be solved via saddle-point approximation
and gives a gaussian-like distribution \emph{(in blue)}. Around $V=\overline V$ \emph{(gray region)} our approximations 
for the solution of the differential equation break down. 
At larger volumes the contour of integration can be deformed around the cut. 
As long as $V\lesssim V_\epsilon$ the integral is dominated by a region that is much larger than the distance $z_{cut}$ 
between the beginning of the cut and the origin, the integral is thus equivalent to the integral
over a cut that starts at $z=0$ \emph{(dotted red contour)} giving a power-law behavior for $\rho(V,\tau)~ \emph{(in light red)}$.
At $V\gtrsim V_\epsilon$, the contour integral ``sees'' the distance $z_{cut}$  between the beginning of the cut and the origin,
and it produces an exponential tail \emph{(in dark red)}. \newline
\mbox{}~~~ Second case $\sqrt\epsilon>\pi/{(2\tau)}$ \emph{(second row)}: this time the saddle point works for all the volumes;
for $V\lesssim V_\epsilon$ it produces a gaussian-like tail that is converted into an exponential tail for $V\gtrsim V_\epsilon$.}
\end{center}
\end{figure}

There are two interesting cases:  $\overline V\lesssim V_\epsilon$ and $\overline V\gtrsim V_\epsilon$. 
Let us start with the case $\overline V\lesssim V_\epsilon$.  As $V$ reaches $\overline V$, $\tau_0+\tau$ becomes of order one, and we can not trust anymore the gaussian solution. 
For this reason, in exploring the regime $V\gtrsim \overline V$, analogously to what we did in the case of $\Omega\geq 1$, we  close the contour along the cut in the negative real $z$ axis (see fig.~\ref{fig:zrho-Wl1}), and perform the integral of the imaginary part of $f(\tau;z) e^{z V}$
\bea\label{eq:int_cut2}
\rho(V,\tau)=\frac{1}{2\pi i}\int_{z_{\rm cut}}^{+\infty}d |z|\; 2 i\, {\rm Im}[f(\tau;-|z|)] e^{- V |z|}\ .
\eea
Because of the exponential suppression, the integral is dominated by $|z|\sim 1/V>|z_{\rm cut}|$ for $V< V_\epsilon$. In this regime, $|z|\gtrsim |z_{\rm cut}|$,  and from eq.~(\ref{eq:solt0}) we have $\tau_0\sim\log( z/\sigma)$. The condition $|z|\sim 1/V$ amounts to ${\rm Re}[\tau_0]\sim -\log V$ (notice that ${\rm Im}[\tau_0]\simeq i\pi$). Therefore in this regime $\tau+{\rm Re}[\tau_0]\sim \log(\overline V/V)\lesssim -1$, and we can use the linearized approximation $f_{\rm lin}(\tau;z)$ for  $f(\tau;z)$. The imaginary part of $f_{\rm lin}(\tau;z)$ on the negative $z$ axis is rather complicated. However, we can approximate $\cos(\sqrt{\epsilon}(\tau+\tau_0))$ with a constant of order one. Then, the integral (\ref{eq:int_cut2}) can be estimated as
\bea\label{eq:int_cut3}
\rho(V,\tau)\sim \int_{z_{\rm cut}}^{+\infty}d |z|\; \, e^\tau z\, e^{- V |z|}\sim \frac{\overline V}{V^2}\ , \quad\quad \overline V\lesssim V\lesssim V_\epsilon \ ,
\eea
where  we ignored constants of order one. We see that in this interval of volumes $\rho(V,\tau)$ decreases as $1/V^2$, exactly matching the analogous regime we found for $\Omega\gtrsim 1$. 
 
As $V$ keeps increasing and becomes larger than $V_\epsilon$, the solution to the boundary condition eq.~(\ref{eq:solt0}), with $z\sim 1/V$, gives $\tau_0\simeq-\pi/(2\sqrt{\epsilon})= -\log V_\epsilon$. In this case, the integral on the discontinuity becomes dominated by the beginning of the cut, and we obtain
\bea\label{eq:int_cut4}
\rho(V,\tau)\sim \int_{z_{\rm cut}}^{+\infty}d |z|\; \, e^\tau \frac{1}{V_\epsilon}\, e^{- V |z|}\sim \frac{\overline V}{V_\epsilon}\frac{1}{V}e^{-\frac{\sigma}{e}\sqrt{\epsilon}V/V_\epsilon}\ , \quad\quad \overline V\lesssim V_\epsilon\lesssim V \ ,
\eea
where again we have ignored constants of order one, and where we have used that $z_{\rm cut}\simeq -\sigma\sqrt{\epsilon}/(eV_\epsilon)$. For volumes larger than $V_\epsilon$, $\rho(V,\tau)$ decreases exponentially (this exponential tail was also recently found in  \cite{Winitzki:2008ph}).

Notice how the two solutions (\ref{eq:int_cut3}) and (\ref{eq:int_cut4}) glue together: for $V\lesssim V_\epsilon$, $\tau_0$ decreases like 
$-\log V$ giving a $1/V^2$ behavior to $\rho(V)$; 
when the volume reaches $\sim V_\epsilon$, $\tau_0$ freezes at a value $\sim (-\log V_\epsilon)=-\pi/(2\eh)$ and the exponential tail ($e^{-\frac{\sigma}{e}\eh\;V/V_\epsilon}$) kicks in. Notice also that the point, where the exponential tail enters, goes to infinity for $\epsilon\to0$ smoothly
matching the result for $\Omega\ge1$.  

Let us now concentrate on the opposite regime. Namely,  as $\epsilon$ increases leaving $\tau$ fixed, at some point $V_\epsilon$ becomes smaller that $\overline V$. In this case, for $V\lesssim V_\epsilon$, we still have the result of eq.~(\ref{eq:fsol1}). However, for larger $V$, $\tau_0$ does not continue to decrease as $-\log V$, because in this case the denominator of eq.~(\ref{eq:Vt0}) goes to zero and determines the behavior of $\tau_0$.  In particular we can expand the denominator of eq.~(\ref{eq:Vt0}) around $\tau_0=-\pi/(2\eh)$
\be \label{eq:dent0}
2\sigma\l[\cos(\eh \tau_0)-\eh\sin(\eh \tau_0)\r]
\approx 2\sigma \eh \l(\tau_0+\frac{\pi}{2\eh}+1 \r)\;,
\ee
that indeed vanishes  for $\tau_0=-\frac{\pi}{2\eh}-1$.
We can now substitute eq.~(\ref{eq:dent0}) into eq.~(\ref{eq:Vt0}) and get
\bea\label{eq:tau0_2}
\tau_0(z_0)&\approx&-\frac{\pi}{2\eh}-1+\frac{e\l(\tau-\frac\pi{2\eh}\r)}{2\sigma\eh}\frac{V_\epsilon}{V}\,, 
\eea
where the third term in the expression for $\tau_0$ is small for $V\gg V_\epsilon$.  We see that $\tau_0$ approaches asymptotically from the positive side the value $\tau_0=-\frac{\pi}{2\eh}-1$. In this regime, $\tau+\tau_0$ is always larger than one and positive, and the gaussian approximation to our solution holds. Further, by substituting  (\ref{eq:tau0_2}) in (\ref{eq:Vt0}), it is straightforward to see that $z_0$ moves on the real axis and approaches $z_{\rm cut}$ from the positive side, reaching $z_{\rm cut}$ only as $V\rightarrow \infty$. This tells us that for $V_\epsilon\lesssim \overline V$, unlike in all the former cases, the saddle point approximation holds for all $V$'s. Plugging in the expression for the saddle point integral we get
\be\label{eq:fsol2}
\rho(V,\tau)\approx {\cal N} e^{-\frac14\l(\tau-\frac\pi{2\eh}\r)^2-\frac\sigma{e} \eh\; V/V_\epsilon}\, , 
\qquad 
\qquad V\gtrsim V_\epsilon\, ,
\ee
with
\be
{\cal N}=\frac{e^{3/2}}
{\sqrt{8\pi\sigma\eh}}\l(\frac{V_\epsilon}V\r)^{3/2}\frac1{V_\epsilon}\l|
\log\l(\frac{\overline V}{V_\epsilon}\r)\r|^{3/2}\,.
\ee
We still have an exponential tail, with the same behavior as we found in the case in which $\overline V$ was smaller than $V_\epsilon$. This is a very intuitive result: as $\epsilon$ increases, the exponential tails becomes relevant at smaller and smaller volumes, and it eats away the main part of the distribution (see fig.~\ref{fig:zrho-Wl1}). This also offers a consistency check between the two ways in which we are computing the inverse Laplace-transform: the saddle point approximation and the integral along the cut.

Finally the normalization of $\rho(V,\tau)$ changes smoothly when crossing the critical point $\Omega=1$
\be
P_{\rm ext}=f(\tau;0)\approx 1-\sigma \eh\; \tau\, e^{\tau-\frac\pi{2\eh}}
=1-6\sigma \eh N_c \frac{e^{6 N_c}}{V_\epsilon}\ .
\ee
In the limit $\epsilon\to0$ the normalization goes to one, but when the exponential tail 
starts to remove the bulk of $\rho(V,\tau)$ (i.e. $V_\epsilon\sim \la V\ra|_{\Omega=1}=e^{6N_c}$)
the volume normalization quickly drops to 0.


For $\epsilon={\cal O}(1)$ our approximations break down and the calculation becomes more complicated: the bound on $\tau_0$ in eq.~(\ref{eq:bt0}) becomes of ${\cal{O}}(1)$ and both the saddle-point approximation and the integral along the cut
are dominated by the region of $z$ where the linearized approximation no longer applies.  However we can still 
say something about the normalization of the volume distribution $P_{\rm ext}$.
For large $\tau$ the solution will still be in the gaussian regime, and we can write $f(\tau;z)$ as
\be
f(\tau;z)=k(\Omega,z) e^{-\frac{(\tau+\tau_0(\Omega,z))^2}{4}}\ ,
\ee
where $k$ and $\tau_0$ are unknown constants that depend on $\Omega$ and $z$. 
From eq.~(\ref{eq:PV}) we know that
\be
P_{\rm ext}=\int_0^\infty\rho(V)dV=f(\tau;0)=k(\Omega,0) e^{-\frac{(\tau+\tau_0(\Omega,0))^2}{4}}\ .
\ee
This formula does not tell us the explicit dependence on $\Omega$ of $P_{\rm ext}$ but since $\tau=6 N_c \sqrt\Omega$,
we can extract the dependence of $P_{\rm ext}$ on the classical number of $e$-foldings $N_c$ (equivalent to the initial position of the inflaton). For large $N_c$ we get
\be
P_{\rm ext} \sim e^{-\Omega(3N_c)^2}=e^{-\Omega (\log V_c)^2}\ .
\ee
The probability not to eternally inflate when $\Omega<1$ goes to 0 exponentially with the square 
of the classical number of $e$-foldings. 

Finally, note that the exponential behavior of $\rho(V)$ at $V>V_\epsilon$  implies that the moments of the distribution 
$\rho(V)$ do not diverge at $\Omega<1$. Naively, this disagree with the results of \cite{Creminelli:2008es}.
 However,  there is no conflict. The point is that here we calculate moments using 
 the probability distribution $\rho(V)$ obtained by taking the infinite time limit. At $\Omega<1$ its normalization is less than one, indicating that there is  a contribution localized 
 at infinity that is not taken into account (cf. with the discussion of the two-site example in section~\ref{sec:2-site}). On the other hand, in the calculation of \cite{Creminelli:2008es} one first finds the moments and only then takes
 the infinite time limit. The latter procedure effectively takes into account the contribution at infinity producing diverging moments at $\Omega<1$.

\subsection{$\Omega=0$: Deeply inside eternal inflation}
We are finally led to study the extreme limit of eternal inflation: the case $\Omega=0$, which corresponds to a completely flat inflationary potential. 

The job is quite easy in this case: the differential equation for $f$
simplifies and we can find an explicit solution. With vanishing $\Omega$ the solution to eq.~(\ref{eq:diffeq}) that stops on the top of the potential in an infinite time reads
\be
f(\tau;z)=e^{\frac12-\frac14\l(\tau+\sqrt{2+4z}\r)^2}\,.
\ee
Then the probability distribution  is
\be
\rho(V,\tau)=\frac{1}{2\pi}\int_{0^+-i\infty}^{0^++i\infty} dz\, e^{\frac12-\frac14\l(\tau+\sqrt{2+4z}\r)^2+zV}\,,
\ee
and to evaluate this integral we can use the saddle point method as before (actually, since the integral is gaussian in $\sqrt{2+4z}$, this procedure is exact).

The saddle point condition in this case reads
\bea
&&S(z)=-\frac12+\frac14\l(\tau+\sqrt{2+4z}\r)^2-zV\,,\quad\\
 &&\Rightarrow \quad S'(z_0)=-V+\l(\frac{\tau}{\sqrt{2+4z_0}}+1\r)=0\,,  \quad \\
 &&\Rightarrow \quad z_0=\frac{\tau^2}{4(V-1)^2}-\frac12\,,
\eea
so that
\bea
S(z_0)&=&\frac{V}{(V-1)}\,\frac{\tau^2}{4}+\frac{V-1}{2}\,, \\
S''(q_0)&=&\frac{2(1-V)^3}{\tau^2}\,, 
\eea
and the expression for the probability distribution is
\be \label{eq:rhoOm0}
\rho(V,\tau)=\frac{\tau}{\sqrt{4\pi}(V-1)^{3/2}}e^{-\frac{V-1}{2}-\frac{V}{(V-1)}\,\frac{\tau^2}{4}}\,.
\ee
The tail of the distribution is again exponential, and this nicely matches the $\Omega\leq1$ case we studied before. 

We can easily also calculate the normalization for $\rho(V,\tau)$,
\be
\int_0^\infty dV\rho(V,\tau)=e^{-\frac{\tau^2}{4}-\frac{\tau}{\sqrt2}}=f(\tau;0)\,,
\ee
which matches with the approximate formula of the previous section.
As long as $\Omega\ll1$ the corrections from the friction term in the differential equation
are small, so that eq.~(\ref{eq:rhoOm0}) is a good approximation in this regime.

%

\subsection{Realistic models: finite barrier effects and slow roll corrections} \label{sec:barrier}
So far, we  worked in the approximation of an infinitely long  inflaton potential and treated
 $\Omega$ and $H$ as constants.  In a realistic situation both these assumptions do not hold:
 $\Omega$ and $H$  change slowly as functions of the inflaton field and the latter 
 can vary only in a finite range. This may be, for example, a consequence of either quantum gravity effects,
 if the potential grows monotonically up to high values of the field, or a reheating region if the potential
 has a maximum, or the steepening of the potential itself at a certain region.
 Let us discuss at the qualitative level how these effects change our results.
 
 Let us begin with the consequences of the finite range of the inflaton field. The 
 details of the underlying physical mechanism are not relevant for our qualitative discussion, so we introduce this effect by including a reflecting barrier in the stochastic process at large values of the inflaton field. In terms of the mechanical problem this implies that we are now looking for a solution that stops at a finite time $\tau_b$ (see (\ref{eq:bc2})).

 The presence of a barrier affects our results in two different ways: it changes the tail of the probability distribution and it slightly 
 shifts the critical value $\Omega_c$ for the transition  to the eternal regime.  Let us start with the first effect. 
 In the non-eternal regime, we have seen that the probability distribution is peaked around a volume of  order $e^{3{\overline N}}$, where ${\overline N}$ is given by (\ref{Nav}) and changes between $N_c$ for large $\Omega$ 
 and $2N_c$ at the transition point $\Omega=1$. Given the relation
 $N_c=\tau/(6\sqrt{\Omega})$ this implies that the typical trajectory undergoes field excursions 
 at most of order $\tau$ up the inflaton potential. 
 Hence putting a barrier at $\tau_b$ does not affect the bulk of the trajectories, and therefore the probability distribution as long as the barrier is far enough from the starting point $\tau\ll \tau_b$. Still,  the barrier cuts the trajectories that would have otherwise crossed it, and therefore we expect an additional  suppression of the tail of the  distribution for volumes corresponding to $\tau\gtrsim \tau_b$, that is for $N\gtrsim N_b$, where $N_b\equiv\tau_b/(6\sqrt{\Omega})$. 
 
As a result, for $N>N_b$ the probability distribution becomes exponentially suppressed as a function of the
volume as opposed to the $V^{-1-{\omega_+/\omega_-}}$ behavior that we found. 
There are several ways to see that the barrier indeed leads to the exponential suppression. For instance,
as we saw in section~\ref{sec:moments}, the function $f(z)$ is analytic at $z=0$ when the barrier is present. Consequently, the whole integration contour for the Laplace transform can be deformed in the region with ${\rm Re}(z)<0$, and at the large volumes
the integral in (\ref{eq:rhoV}) decays as  $e^{-{\rm Re}(z_{cut}) V}$, where $z_{cut}$ is the singularity of the function $f$ at the smallest distance from the real axis.

In fact,
 recently an exponential tail of the volume distribution in the eternal regime
 was calculated in \cite{Winitzki:2008ph}.
 By extending the analysis of 
\cite{Winitzki:2008ph} into the non-eternal regime we verified that, as a result, the exponential suppression
 sets in for a number of $e$-foldings  of order ${N_b}$. We do not provide details of this analysis here, as it would require an extensive introduction into methods of \cite{Winitzki:2008ph} and would take us too far away from the main line of our paper (instead, in the concluding section~\ref{conclusions}, we will present an intuitive argument explaining
 the origin of this exponential behavior).
This also  agrees with what we found in the two site models, where the probability distribution decreases exponentially at large volumes both in the eternal and the non-eternal regime.

Notice that, since $\Omega>1$ implies $N_c<S_{dS}/12$ (see eq.~(\ref{vague_limit1/4})), in the same regime we also have an upper bound
on $N_b$, which has to satisfy the same condition $N_b<S_{dS}/12$. This also means that the behavior of $\rho(V)$ for
volumes larger than $e^{S_{dS}/2}$ is always exponential in the volume, \emph{i.e.} $\sim e^{-{\rm const} V}$.
This property of the probability distribution will be further discussed in the next section 
in connection with the bound on the number of $e$-foldings.
 
 The second effect of the barrier is related to the first. The presence of the barrier cuts out some of the trajectories 
producing  the largest volumes and thus may slightly delay  the entrance in the eternal regime   to $\Omega_c<1$. This shift can be calculated in the following way. Since $\Omega_c$ is defined as the value of $\Omega$ where $P_{\rm ext}$ starts deviating from one, and since $P_{\rm ext}=f(\tau;0)$, we need to study the solution $f$ for $z\rightarrow 0$.
Consider some fixed $\Omega<1$, and let us recall that in this case the oscillator is not anti-over-damped (eq.~(\ref{eq:diffeq})). Around $f=1$, we can linearize the potential and obtain an oscillating solution with a period 
\be
T=\frac{2\pi}{\sqrt{1-\Omega}} \,,
\ee
independently of the initial velocity, as the oscillations are harmonic (see the dotted lines of fig.~\ref{fig:fO}b). 
Consequently, in the linear regime it takes an amount of time equal to $T/4$ for the solution to come at rest.
The returning force for the actual unharmonic potential (\ref{eq:Uf}) is smaller than for the corresponding
harmonic potential, so it always takes longer than $T/4$ to come at rest (and can take arbitrarily long, as at $\Omega<1$
there exists a solution that stops at the top of the hill in an infinite time). Consequently, if the barrier is close enough,
$\tau_b<T/4$, there is no solution with the appropriate boundary condition, $\dot{f}(\tau_b;0)=0$ (except the trivial solution $f(\tau;0)=1$), even at $\Omega<1$
and inflation is not eternal.
This argument implies that the  critical value for $\Omega$ is determined by
\be
\tau_b=\frac\pi{2\sqrt{1-\Omega_c}} \quad \Rightarrow \quad \Omega_c=1-\l(\frac{\pi}{2\tau_b}\r)^2\,.
\ee

Note that for $\Omega\lesssim\Omega_c$, the volume where the barrier effect sets in and changes the tail of the distribution is of order $V\lesssim V_b^2= e^{6N_b}=e^{\pi/2\sqrt{1-\Omega_c}}$, that is always exponentially larger
then the volume $V_\epsilon$, where the exponential suppression found in section~\ref{sec:omegal1}
sets in. Consequently, in the eternal regime the barrier affects only the tail of the distribution that is already suppressed
as $e^{-{\rm const} V}$.

Let us now discuss how the dependence of $\Omega$ and $H$ on the inflaton field changes our results.
Let us see first when this dependence become important. Throughout this paper we have been interested in what happens
close to the eternal regime in the limit $M_{\rm Pl}\gg H$. In other words we have taken the limit $H/M_{\rm Pl}\to 0$,
while keeping $\Omega$ fixed. This limit implies that we have been working in the extreme slow roll regime. 
Indeed, in this limit one has,
\[
{\dot H\over H^2}\sim\Omega {H^2\over M_{\rm Pl}^2}\to 0\;.
\]
However, this does not imply that one can completely neglect the field dependence of $\Omega$ and $H$ 
as we have another large parameter---the variation of the inflaton field or, equivalently, the number of $e$-foldings.
For instance, by taking the variation of the Friedmann equation, one has
\[
M_{\rm Pl}^2{H\Delta H}\sim V'\Delta\phi
\]
and the condition that $H$ can be treated as a constant reads
\be
\label{dhh}
{\Delta H\over H}\sim {V'\Delta\phi\over M_{\rm Pl}^2 H^2}\sim \Omega{H^2\over M_{\rm Pl}^2}N_{c}\lesssim 1\;.
\ee
So all our results apply if one takes the limit $H/M_{\rm Pl}\to 0$ while keeping  $\Omega$ fixed
 and $N_c$ small enough such that the inequality (\ref{dhh}) holds. Note that (\ref{dhh}) does not prevent us
 from considering arbitrarily large $N_c$ if $H/M_{\rm Pl}$ is taken to be sufficiently small. 
 
 Nevertheless, one may wonder what happens to the shape of the volume probability distribution
 for longer inflaton trajectories, such that the variation of $H$ (and $\Omega$) has to be taken into account.
 In this case the coefficients in front of the different terms in the mechanical equation acquire a $\phi$ dependence
 \cite{Winitzki:2001np}. This clearly may affect the details of the shape of the probability distribution.
 However, as long as $H\ll M_{\rm Pl}$  we still expect these effects to be small,
 with $\rho(V)$ still sharply  peaked around the average number of $e$-foldings $\overline N$, which takes values between $N_c$ and $2N_c$.
 One can check this statement by using the techniques of section~\ref{sec:moments} to calculate the average
 and the higher moments of the distribution. We can consider, for instance, the linear differential equation (\ref{eq:diffav})
 for the average in the generic case, where $\Omega$ and $H$ can be functions of the inflaton field $\phi$ 
 (or equivalently of $\tau$). In the variable $\tau$, defined now as
 \be
 \tau \equiv 6 \int \sqrt{\Omega} d N_c\,,
 \ee
 the differential equation for the average is the same as in eq.~(\ref{eq:diffav}) with a small correction (of order $H^2/M_{\rm Pl}^2$) 
 to the anti-friction term, which now depends on $\tau$ implicitly via $\Omega$. Since in the non-eternal regime
  the anti-friction term is minimized at $\Omega=1$, 
  by setting $\Omega=1$ we would obtain a trajectory that has a faster 
   velocity at each moment of time in order to stop at the same moment as required by the boundary condition
    (\ref{eq:bcav}).
 It follows that  the average is always smaller than $e^{\tau}$, which, from eq.~(\ref{precise_classical}), is smaller than $e^{S_{dS}/2}$. This implies that the bound $\overline N<S_{dS}/6$ still holds.
 
 In fact, we can also find an approximate expression for the average volume in the general case.
 Namely, for non-constant $\Omega$, we can replace (\ref{eq:Vav1}) by
\be 
\lim_{\tau_b\to \infty} \la V\ra = e^{\int\omega_- d\tau},
\ee
 which is a good approximate solution as soon as $\d_\tau\Omega\ll\Omega$. 
 This gives for the average number of e-foldings
 %
 \be
 3\overline N
 =\int \omega_- d\tau\,.
 \ee
 Analogous arguments can be applied for higher moments; as a result we see that the distribution
  remains peaked around the average value that takes value between $N_c$ and $2N_c$.

 \section{Discussion}
\label{conclusions}

To summarize, in this paper we calculated explicitly the probability distribution $\rho(N)$
for the volume of the Universe after a period of slow-roll inflation (as before, by the number of $e$-foldings $N$ 
we understand  one third of the logarithm of the total volume produced during inflation, $N={1\over 3}\log V$). 
Our results cover both the eternal and the non-eternal regime.
Let us start this concluding section by summarizing these results and then by explaining how all different 
kinds of behavior we found for $\rho(N)$ have an intuitive physical explanation. 

In general, the function $\rho(N)$ exhibits three qualitatively different regions. 
Namely, in the non-eternal regime $\Omega>1$ it is peaked at $N\sim {\overline N}$, where the average
number of $e$-foldings ${\overline N}$ is given by
\be
\label{sumNav} 
{\overline N}={2N_c\over 1+\sqrt{1-\Omega^{-1}}}\;.
\ee
For $N\lesssim {\overline N}$ it has a gaussian form,
\be
\label{sumgaussian}
\rho(N)\propto e^{-{(3N-3{\overline N})^2\over 2\sigma^2}}\;,
\ee
with a width $\sigma$ given by
 \begin{eqnarray}
 \label{sumwidthNav}
  \sigma^2={2\over\Omega(1+\sqrt{1-\Omega^{-1}})^2}\;.
 \end{eqnarray}
The behavior changes at $N \gtrsim {\overline N}$ where the probability distribution becomes exponential in $N$ (or, equivalently,
power-law in the volume $V$),
\be
\label{sumexp}
\rho(N)\propto e^{-6\Omega N\l(1+\sqrt{1-{1\over \Omega}}\r)}=V^{-2\Omega \l(1+\sqrt{1-{1\over \Omega}}\r)}\;.
\ee
Finally, if a barrier preventing the inflaton to take arbitrary large values is present, this power-law 
tail becomes further suppressed at larger volumes and turns into an exponential in the  {\it volume} 
\be
\label{sumexpexp}
\rho(N)\propto e^{-c e^{3N}}=e^{-c V}\;.
\ee
This behavior sets in at $N\sim N_b$, where $N_b$ is the number of $e$-foldings on the classical inflaton trajectory from the barrier  till reheating.

What changes in the eternal regime, $\Omega<1$, is that
the exponential behavior (\ref{sumexpexp}) sets in at $N\simeq{\pi/ 6 \sqrt{1-\Omega}}$
even if the barrier is absent. If  $(1-\Omega)$ is not too small 
this happens at $N<{\overline N}$ so that the 
gaussian
regime (\ref{sumgaussian}) interpolates directly to the superexponential one
(\ref{sumexpexp}) without intermediate exponential behavior (\ref{sumexp}).

It is rather straightforward to understand the origin of the above three different
types of behavior for $\rho(N)$ at the intuitive level. First, 
to produce
a small number of $e$-foldings $N\ll N_{c}$, the inflaton field during the first few $e$-foldings 
needs to perform a jump in the whole volume to a value of $\phi$ corresponding to
an average number of $e$-folding equal to $N$.
The  inflaton fluctuations away from the classical trajectory follow, at early times, a gaussian distribution 
(because of the random walk) and the size of
the jump in field space is directly proportional to $(N-{\overline N})$, so the probability of such a jump
is suppressed by $e^{-c (N-{\overline N})^2}$ ($c$ is a constant) in agreement with our result (\ref{sumgaussian}) at small volumes.

On the other hand, the least expensive way to produce a large number of $e$-foldings $N\gg N_{c}$ is for one Hubble patch to go high up the potential, till values of the field corresponding to have a classical trajectory with $N$ $e$-foldings.
If such a fluctuation happened, the probability to produce at least $N$ $e$-foldings becomes of order one. To estimate the probability
of such a fluctuation, note again that the probability distribution for the inflaton fluctuations $\Delta\phi$ around the
classical trajectory is gaussian. The crucial difference with the small volume case is that now  it is not necessary for the fluctuation
to happen in a short period of time, and the variance of the inflaton distribution grows linearly as a function of time $t$.
As a result, the probability $p$ of the fluctuation is maximized for times $t$ corresponding to order $N$ $e$-foldings, and
depends exponentially on $N$, $p\propto e^{-c_1
\Delta\phi^2/t}\propto e^{-c_2 N}$ in agreement  with (\ref{sumexp}). This argument is essentially identical to the one
provided in \cite{Creminelli:2008es} to explain why high enough moments of the volume distribution diverge in the non-eternal regime if no barrier is introduced.

The last argument fails at sufficiently large volumes both in the presence of a barrier at high values of the inflaton field
and in the eternal regime. In the first case it fails because there is a limit on the maximum length of the classical trajectory, while in the second because even if a fluctuation to high values of the inflaton field happened, one is not guaranteed 
to end up with a {\it finite} number of $e$-foldings (in fact, the higher the fluctuation is the smaller is the probability for inflation to terminate). In both cases one ends up with  getting a superexponential suppression of the probability distribution at large volumes, (\ref{sumexpexp}). This can roughly be explained by the necessity for an exponential 
number of independent and rather improbable events (corresponding to the individual Hubble volumes produced during inflation) to happen. Namely, in the eternal regime, inflation should terminate for $\sim e^{3N
}$ Hubble patches and all their children (while in the non-eternal regime the probability for this  to happen is equal to one). In the non-eternal case with a barrier, for $N\gg N_b$, an order $\sim e^{3N
}$  Hubble patches have to live an anomalously long time, corresponding to a number of $e$-foldings much larger than $N_b$.

To summarize, we see that our results on the shape of the probability distribution of the reheating volume
not only pass a number of consistency checks with results obtained by different methods, but also 
can be understood in a rather intuitive level.
We believe making the explicit form of the volume probability available  may help to understand the geometry of eternal inflation and offers also a natural
``theoretical" observable that can be useful in further studies of eternal inflation. 

A further motivation for doing our calculation was to establish whether a quantum version of the bound 
\be \label{vague_limit_2}
3N \leq cS_{dS}\;,
\ee
exists and, if so, whether this bound can be made sharp, 
{\it i.e.} whether there is a concrete value for the coefficient $c$ in (\ref{vague_limit_2}). 
Our results provide a positive answer to both questions. Namely, we find
that the quantum version of the
bound (\ref{vague_limit_2})  can be formulated as follows:

\emph{The probability for slow-roll inflation to produce a finite volume larger than $e^{S_{dS}/2}$, where $S_{dS}$ is de~Sitter
entropy at the end of the inflationary stage, 
is suppressed below
 the uncertainty due to non-perturbative quantum gravity
effects.}

 Indeed, our results imply
that an inflaton trajectory  with more than $2N_c$ $e$-foldings and such that inflation terminates globally in the entire space is
exponentially unprobable, $\sim e^{-N}$. 
Given that in the non-eternal regime we have the bound (\ref{vague_limit1/4}), we conclude
 that the probability
for slow-roll inflation to last more than $S_{dS}/6$ $e$-foldings and to terminate globally in the entire space
is smaller than the uncertainty coming from non-perturbative quantum gravitational effects ($\sim e^{-S_{\rm dS}}$). 
In fact, there is an even stronger statement
that follows from the observation of section~\ref{sec:barrier} 
about the bound on the barrier position $N_b$. Indeed, as long as $\Omega\gtrsim 1$, $N_b$ is smaller  than $S_{dS}/12$, 
which implies that the transition to the regime where
 $\rho(V)$  drops exponentially with the volume ($e^{-{\rm const}\cdot V}$) starts exactly when the
  produced volume  starts violating the bound,
 $V>e^{S_{dS}/2}>e^{6N_b}$.
 Consequently, the probability to produce
 much larger volumes $V\gg e^{S_{dS}/2}$ is super-exponentially suppressed, $\sim e^{-e^{S_{dS}}}$.
In turn this produces a super-exponentially small probability to produce volumes $V\gg e^{S_{dS}/2}$.

We believe this provides a  non-trivial test of the de~Sitter complementarity idea.
However, one may wish to go further and ask whether the particular value $c=1/2$ that we found provides
any insights into how the de~Sitter complementarity works. 
Of course, all our calculations were done in the limit where gravity is non-dynamical, therefore it is not clear whether the 
value $c=1/2$ really tells us something fundamental about the properties of de~Sitter space. Let us however assume
 that it does and speculate on what would be the interpretation of this particular value. 
 Recall that the original motivation for the bound (\ref{vague_limit}) is coming from the idea that 
 the black hole complementarity applies in the de~Sitter case as well, 
 so that the global effective field theory description of the FRW slices breaks down
 and information about the outside observers gets released in de~Sitter fluctuations. 
Then, if inflation ends, there is the danger of violating the linearity of quantum mechanics 
by creating a ``quantum xerox"---one can see the information twice, 
first holographically in the de~Sitter fluctuations and later in a direct way when the corresponding mode comes in.

Recall that the analogous paradox could potentially arise in the black hole case after
one  measures more than $S_{bh}/2$ Hawking quanta ($S_{bh}$ is the black hole entropy).
In this case, the factor $1/2$ arises because, as pointed out by Page \cite{Page:1993wv}, if one measures
$k$ degrees of freedom of a larger system described by $n$ degrees of freedom, generically, 
the resulting density matrix looks thermal and carries less than a single bit of information as long as $k<n/2$. 

This does not give rise however to any paradox: indeed, if  one waits a long enough time outside the black hole, so that $\sim S_{bh}/2$ Hawking quanta are emitted, 
it is impossible to observe the same information a second time inside the black hole, 
since it gets destroyed by the curvature singularity.

It is tempting to interprete our result by saying that the de~Sitter analogue of 
one Hawking quanta is produced every time a new Hubble patch is created from the volume where the observer
lives ({\it{i.e}} every $1/3$ of $e$-folding). 
Then the quantum-xerox problem would arise after $S_{dS}/6$ $e$-foldings but it does not
because of the bound. 

Although  it may sound quite natural, this interpretation raises a number of issues. 
First, as argued in \cite{Parikh:2008iu}, a single observer in {\it eternal} de~Sitter can never run into the xerox paradox, 
as the largest amount of entropy that can be stored within a single causal patch is bounded by the Bekenstein area 
of the largest black hole that fits in a single patch. The latter is equal to $S_{dS}/3$ and smaller than the minimal amount 
($S_{dS}/2$) required to extract the first bit of information.  

However we are in a rather different situation here, as inflation eventually terminates. 
Therefore should the bound be not there, one would have the possibility to measure
 $S_{dS}/4$ quanta, give them to a friend who leaves for a different Hubble patch, then measure other $S_{dS}/4$
 quanta and, after  inflation ends, the two could  meet  and in this way collect $S_{dS}/2$ quanta in total.
 Then the bound suggests that for the purpose of extracting information it is legitimate to collect
 quanta in one Hubble patch and keep them until the end of inflation in another.
 However, this reasoning also implies that it should not be possible to use quanta 
 collected in {\it different} Hubble patches for extracting information---otherwise one would be 
 forced to conclude that the actual bound on the number of $e$-folding is much stronger, 
 $N\lesssim \log{S}$.
 
 Notice also that if inflation does not terminate everywhere, then there is no problem 
 of duplication of information if one observer lasts in the inflationary phase for longer than $S_{dS}/6$ $e$-foldings: 
 when the observer undergoes reheating and enters the Minkowski phase she is never able to see 
 all the rest of the de~Sitter space.

Another puzzle with drawing the parallel between the factor 1/2 in our bound and the one in the Page argument
is that, if there are $n$ light species around then, likely, the rate of how fast the information gets released 
is proportional to the number of species (at least this is so in the black hole case). 
On the other hand our bound does not depend on $n$ 
(it could if the light species are scalars but does not for other fields).

Finally, it is worth mentioning yet another reason why one should be cautious in taking the value $c=1/2$ too seriously.
Namely, here we focused on slow-roll models of inflation.  
In the effective field theory language of \cite{Cheung:2007st} these correspond to the models 
where the inflaton perturbations have sound velocity $c_s$ equal to one.
The classical version of the bound (\ref{vague_limit}) was proven in \cite{ArkaniHamed:2007ky} for  general inflationary models, 
including those with small $c_s$. 
In fact at small $c_s$ the classical bound becomes even stronger, $N_c\lesssim c_s^5 S_{dS}$.
However the reason it arises is somewhat different. 
At small sound velocities the strong coupling regime, where the effective field theory breaks down, 
sets in always before the eternal inflation regime. So the bound (\ref{vague_limit}) in this case follows 
from the requirement that the system is weakly coupled and the null energy condition is not violated.
Consequently, at the present stage we cannot exclude the possibility of violating the bound in the strongly coupled
regime, although this possibility appears highly unlikely.

To summarize, we proved the quantum version of the bound on how long slow-roll inflation can last
without becoming eternal. The existence of such a bound (eq.~(\ref{vague_limit_2})) provides 
non-trivial support to de~Sitter complementarity ideas, while it is still unclear whether the value of the coefficient 
 $c=1/2$ appearing in the bound that we found has a definite physical meaning.
Before finishing, it is worth stressing that we believe that, independently of the answer to the last question, 
explicit calculations providing a detailed quantitative understanding of the transition to the eternal inflation regime, 
like those we preformed in the current paper, have an independent value. 
Indeed,  it appears to us that a detailed understanding of the geometry and the dynamics of
the eternally inflating Universe might be an important step to reach a final verdict 
on the puzzling issues raised by the observation of the cosmic acceleration and the string landscape.


\section*{Acknowledgments}

We would like to thank Paolo Creminelli for initial collaboration in this project. 
It is a pleasure to thank Nima Arkani-Hamed, Raphael Bousso, Willy Fischler, Ben Freivogel, Alan Guth, Shamit Kachru, Andrei Linde, Hong Liu, Juan Maldacena, Alberto Nicolis, Sonia Paban, Uros Seljak, Steve Shenker,  Eva Silverstein, Lenny Susskind, Enrico Trincherini, Alex Vilenkin and Matias Zaldarriaga for stimulating discussions. The work of SD and LS is supported in part respectively by the National  Science Foundation under Grants No. PHY-0455649 and PHY-0503584.

\appendix
\section{Volume average from the inflaton stochastic equation \label{sec:avvol}}
As a cross-check of the method used in the paper, derived from the bacteria model,
in this appendix we present an alternative calculation of the average volume $\la V\ra$
directly from the inflaton equations. The result is an improvement of
a very similar computation performed in \cite{Creminelli:2008es}.
By definition the average of the volume is given by
\be
\label{defVav}
\la V\ra=\int_0^\infty dV\;V\rho(V,\phi)\;,
\ee
where the adimensional variable $V$ is the volume expressed in units of the initial volume. 
This reduces to compute
\be
\label{Vavpr}
\la V\ra=  \int_{t_0}^{\infty} \! dt \, e^{3Ht}p_r(t)\;,
\ee
where $p_r(t)$ is the probability that at a given point $\vec x$ the field reaches the reheating
value $\phi_r$ at time $t$. By translational invariance this probability does not depend on the point $\vec 
x$. Now the problem is reduced to compute $p_r(t)$. The latter is related to the probability $P(\bar\phi,t)$
for the inflaton to have a value $\bar\phi$ at time $t$ by 
\be
\label{reheating_prob}
p_r(t)= -\frac{d}{d t} \int^{+\infty}_{\phi_r=0} \! d\bar\phi  \; P(\bar\phi,t) \; .
\ee
$P(\bar\phi,t)$ can be found by solving the classic stochastic 
diffusion equation \cite{starobinsky,Linde:1986fd,Linde:1986fc}
\begin{eqnarray}\label{eq:stoch.eq}
\partial_{\sigma^2}  {\tilde P}(\psi,\sigma^2) =\frac{1}{2} \partial^2_\psi {\tilde P}(\psi,\sigma^2)\; .
\end{eqnarray}
where ${\tilde P}(\psi,\sigma^2)\equiv P(\bar\phi,t)$ and
\be
\label{phidef}
\psi\equiv\bar\phi-\phi-\dot\phi t
\ee
is a Gaussian field with variance $\sigma^2$ that grows linearly with time  
\be
\label{sigma}
\sigma^2={H^3\over 4\pi^2}t \,.
\ee
$\psi$ represents the fluctuations around the classical motion and undergoes a random walk.
In the case the inflaton lives in a infinitely long potential and the reheating point is at $\phi_r=0$, the solution of 
(\ref{eq:stoch.eq}) is given by
\begin{equation}
{\tilde P}(\psi,\sigma^2)=\frac{1}{\sqrt{2\pi\sigma^2}}\left(e^{-\frac{\psi^2}{2\sigma^2}}-e^{-8\pi^2\frac{\dot\phi\,
\phi}{H^3} }e^{-\frac{\left(\psi+2\phi\right)^2}{2\sigma^2}}\right)\, ,
\end{equation}
which implies
\be
\label{eq:prexp}
p_r(t)=\sqrt{2\pi }\frac{\phi}{(H\, t)^{3/2}}e^{-2\pi^2\frac{(\phi+\dot\phi\, t)^2}{H^3 t}} \, .
\ee
We see that the integral in (\ref{Vavpr}) converges only when 
\be
\Omega\equiv\frac{2\pi^2}{3}\frac{\dot\phi^2}{H^4}\geq 1 \ .
\ee 
This is the first signal that the system has entered the eternal inflation regime.
For $\Omega\geq 1$, we can explicitly perform the integral in (\ref{Vavpr}), to obtain:
\be \label{eq:vavinfl}
\langle V \rangle= e^{-2\pi\left(\frac{2\pi\dot\phi+\sqrt{(2\pi\dot\phi)^2-6H^4}}{H^2}\right)\frac{\phi}{H}}=e^
{2\sqrt{6}\pi\left(\sqrt{\Omega}-\sqrt{\Omega-1}\right)\frac{\phi}{H}}=\l\{\begin{array}{ll}V_c\,,& \Omega\gg1\\V_c^2\,,&\Omega\to1\end{array}\r. \ , \quad\quad \Omega\geq 1\ .
\ee
where $$V_c\equiv e^{\sqrt{\frac{6\pi^2}{\Omega}}\frac{\phi}{H}}\,,$$ is the volume in the classical limit.
The result, which agrees with eq.~(\ref{eq:vavdiff}), is plotted in fig.~\ref{fig:Nav}, 
where we can see the singular behavior at $\Omega=1$.


\end{document}